\documentclass[longbibliography,aps,prx,preprint,superscriptaddress]{revtex4-2}
\usepackage{latexsym}
\usepackage{graphicx}
\usepackage{color}
\pdfoutput=1
\begin{document}


\title{Nonlinear and nonreciprocal transport effects in untwinned thin films of ferromagnetic Weyl metal SrRuO$_{3}$}



\affiliation{Institute of Physics, Academia Sinica, Nankang, Taipei 11529, Taiwan}
\affiliation{Department of Physics, National Taiwan University, Taipei 10617, Taiwan}
\affiliation{Department of Physics, Applied Physics and Astronomy, Binghamton University, Binghamton, New York 13902, USA}
\affiliation{Department of Physics, California Institute of Technology, Pasadena, California 91125, USA}
\affiliation{Scientific Research Division, National Synchrotron Radiation Research Center, Hsinchu 30076, Taiwan}
\affiliation{Nano Science and Technology, Taiwan International Graduate Program, Academia Sinica and National Taiwan University, Taipei, Taiwan}
\affiliation{Physics Division, National Center for Theoretical Sciences, Taipei 10617, Taiwan}
\affiliation{Graduate Institute of Photonics and Optoelectronics, National Taiwan University, Taipei, Taiwan}

\author{Uddipta Kar}\thanks{These authors contributed equally to the work.}
\affiliation{Institute of Physics, Academia Sinica, Nankang, Taipei 11529, Taiwan}
\affiliation{Nano Science and Technology, Taiwan International Graduate Program, Academia Sinica and National Taiwan University, Taipei, Taiwan}
\author{Elisha Cho-Hao Lu}\thanks{These authors contributed equally to the work.}
\affiliation{Institute of Physics, Academia Sinica, Nankang, Taipei 11529, Taiwan}
\author{Akhilesh Kr. Singh}\thanks{These authors contributed equally to the work.}
\affiliation{Institute of Physics, Academia Sinica, Nankang, Taipei 11529, Taiwan}
\author{P. V. Sreenivasa Reddy}\thanks{These authors contributed equally to the work.}
\affiliation{Department of Physics, National Taiwan University, Taipei 10617, Taiwan}
\author {Youngjoon Han}
\affiliation{Department of Physics, California Institute of Technology, Pasadena, California 91125, USA}
\author {Xinwei Li}
\affiliation{Department of Physics, California Institute of Technology, Pasadena, California 91125, USA}
\author{Cheng-Tung Cheng}
\affiliation{Institute of Physics, Academia Sinica, Nankang, Taipei 11529, Taiwan}
\author{Song Yang}
\affiliation{Scientific Research Division, National Synchrotron Radiation Research Center, Hsinchu 30076, Taiwan}
\author{Chun-Yen Lin}
\affiliation{Scientific Research Division, National Synchrotron Radiation Research Center, Hsinchu 30076, Taiwan}
\author{I-Chun Cheng}
\affiliation{Graduate Institute of Photonics and Optoelectronics, National Taiwan University, Taipei, Taiwan}
\author{Chia-Hung Hsu}
\affiliation{Scientific Research Division, National Synchrotron Radiation Research Center, Hsinchu 30076, Taiwan}
\author{David Hsieh}
\affiliation{Department of Physics, California Institute of Technology, Pasadena, California 91125, USA}
\author{Wei-Cheng Lee}
\affiliation{Department of Physics, Applied Physics and Astronomy, Binghamton University, Binghamton, New York 13902, USA}
\author{Guang-Yu Guo}\email{gyguo@phys.ntu.edu.tw}
\affiliation{Department of Physics, National Taiwan University, Taipei 10617, Taiwan}
\affiliation{Physics Division, National Center for Theoretical Sciences, Taipei 10617, Taiwan}
\author{Wei-Li Lee}\email{wlee@phys.sinica.edu.tw}
\affiliation{Institute of Physics, Academia Sinica, Nankang, Taipei 11529, Taiwan}


\date{\today}

\begin{abstract}
The identification of distinct charge transport features, deriving from nontrivial bulk band and surface states, has been a challenging subject in the field of topological systems. In topological Dirac and Weyl semimetals, nontrivial conical bands with Fermi-arc surface states give rise to negative longitudinal magnetoresistance due to chiral anomaly effect and unusual thickness dependent quantum oscillation from Weyl-orbit effect, which were demonstrated recently in experiments. In this work, we report the experimental observations of large nonlinear and nonreciprocal transport effects for both longitudinal and transverse channels in an untwinned Weyl metal of SrRuO$_3$ thin film grown on a SrTiO$_{3}$ substrate. From rigorous measurements with bias current applied along various directions with respect to the crystalline principal axes, the magnitude of nonlinear Hall signals from the transverse channel exhibits a simple sin$\alpha$ dependence at low temperatures, where $\alpha$ is the angle between bias current direction and orthorhombic [001]$_{\rm o}$, reaching a maximum when current is along orthorhombic [1\={1}0]$_{\rm o}$. On the contrary, the magnitude of nonlinear and nonreciprocal signals in the longitudinal channel attains a maximum for bias current along [001]$_{\rm o}$, and it vanishes for bias current along [1\={1}0]$_{\rm o}$. The observed $\alpha$-dependent nonlinear and nonreciprocal signals in longitudinal and transverse channels reveal a magnetic Weyl phase with an effective Berry curvature dipole along [1\={1}0]$_{\rm o}$ from surface states, accompanied by 1D chiral edge modes along [001]$_{\rm o}$.             

\end{abstract}


\maketitle

\newpage
\section{Introduction}
Since the first experimental demonstration of a quantized conductance from counter-propagating edge spin channels in HgTe quantum well system \cite{Konig2007}, topological materials have become one of the main research focuses in condensed matter physics and materials science. The two dimensional (2D) quantum spin Hall phase originates from inverted bulk bands that cross near the system's boundary, revealing one dimensional helical edge states and thus the observed conductance quantization, which is also known as the 2D topological insulator (TI) phase and also recently reported in several other 2D systems \cite{Du2015,Fei2017,Tang2017}. Extending to 3D TI, the existence of a nontrivial bulk band topology with an intrinsic topological invariant gives rise to unusual gapless Dirac surface states, which was confirmed in experiments using surface sensitive angle-resolved photoemission spectroscopy and scanning tunneling microscopy \cite{Hsieh2008,Alpi2010,Hasan2010}. More recently, a remarkable advancement was made by the observation of the quantized anomalous Hall conductance at zero magnetic field in a magnetic TI \cite{Chang2013,Kou2014,Checkelsky2014}, and it is a unique transport signature due to the topological nature of the system, which was theoretically predicted long ago \cite{Haldane1988}. 

In topological Dirac and Weyl semimetals (WSM), nontrivial crossings appear in the bulk bands near the Fermi surface \cite{Wan2011,Wang2012}, and charge transport is overwhelmed by the unusual chiral charge excitations near nodal points with Berry phase $\pi$, showing superior electron mobility due to the suppressed backscattering by spin-momentum lock effect \cite{Liang2015,Huang2015} and negative longitudinal magnetoresistance (MR) for aligned electric field and external magnetic field due to the chiral anomaly effect \cite{Xiong2015,Armitage2018}. In addition, unique Fermi-arc surface states \cite{Wan2011,Armitage2018} appear on a surface of a WSM, connecting the projected Weyl-node pair, where a number of intriguing novel charge transport features have been predicted theoretically \cite{Armitage2018,Potter2014,Waw2021}. For a ferromagnetic WSM, there can be a minimum number of one Weyl-node pair with opposite chiral charges near the Fermi surface, accompanied by 1D chiral zero edge modes perpendicular to the connecting momentum of the Weyl-node pair. In this work, we report the experimental observations of nonlinear Hall signals \cite{Gao2014,Sodemann2015,Ma2019} for $T \leq$ 10 K in the untwinned thin film of ferromagnetic Weyl metal SrRuO$_3$ (SRO) grown on a miscut SrTiO$_{3}$ (STO) substrate. Rigorous bias current dependent measurements of the nonlinear Hall signals correspond to an effective Berry curvature dipole (BCD) $\vec{D}$ from surface states along the orthorhombic [1\={1}0]$_{\rm o}$, where the subscript o refers to orthorhombic-phase. Surprisingly, a nonlinear and nonreciprocal transport effect in the longitudinal channel (NRTE) was also observed. It attains a maximum when the bias current is aligned perpendicular to $\vec{D}$, but it becomes vanishingly small when bias current is parallel to $\vec{D}$, which can be attributed to the 1D chiral edge modes as demonstrated previously in the quantum anomalous Hall system \cite{Yasuda2020}. Those results support the intriguing magnetic WSM phase in SRO/STO system with an effective surface $\vec{D}$ along [1\={1}0]$_{\rm o}$ accompanied by 1D chiral edge modes along [001]$_{\rm o}$ that circle around the surface of a SRO thin film.             

\section{Experimental setup}
SRO is known as a ferromagnetic and metallic oxide, showing an orthorhombic crystal structure with $Pbnm$ space group symmetry at room temperature \cite{Koster2012,Kar2021}. In the past, the observed nonmonotonic magnetization (M) dependent anomalous Hall conductivity \cite{Fang2003,Chen2013}, unusual temperature dependent magnon gap \cite{Itoh2016} and softening of the magnon mode at low temperatures \cite{Jenni2019} all pointed to the existence of the Weyl nodes near the Fermi surface, supporting the Weyl metal phase in SRO system. Recently, the advancement in the growth of exceptional quality SRO thin films with ultra-low ruthenium vacancy level was made possible using oxide molecular beam system \cite{Nair2018,Taki2020,Kar2021}. The low residual resistivity at $T$ = 2 K of only about 10 $\mu\Omega$cm for a SRO film with thickness of about 10 nm \cite{Cap2002}, which may largely suppress the smearing of the Weyl nodes due to the rare region effects \cite{Nand2014}, makes it possible to explore various charge transport features associated with the Fermi-arc surface states and Weyl metal phase of SRO in thin film form \cite{Kaneta2022,kar2022}. 

\section{Results}

\begin{figure}
\includegraphics[width=\linewidth]{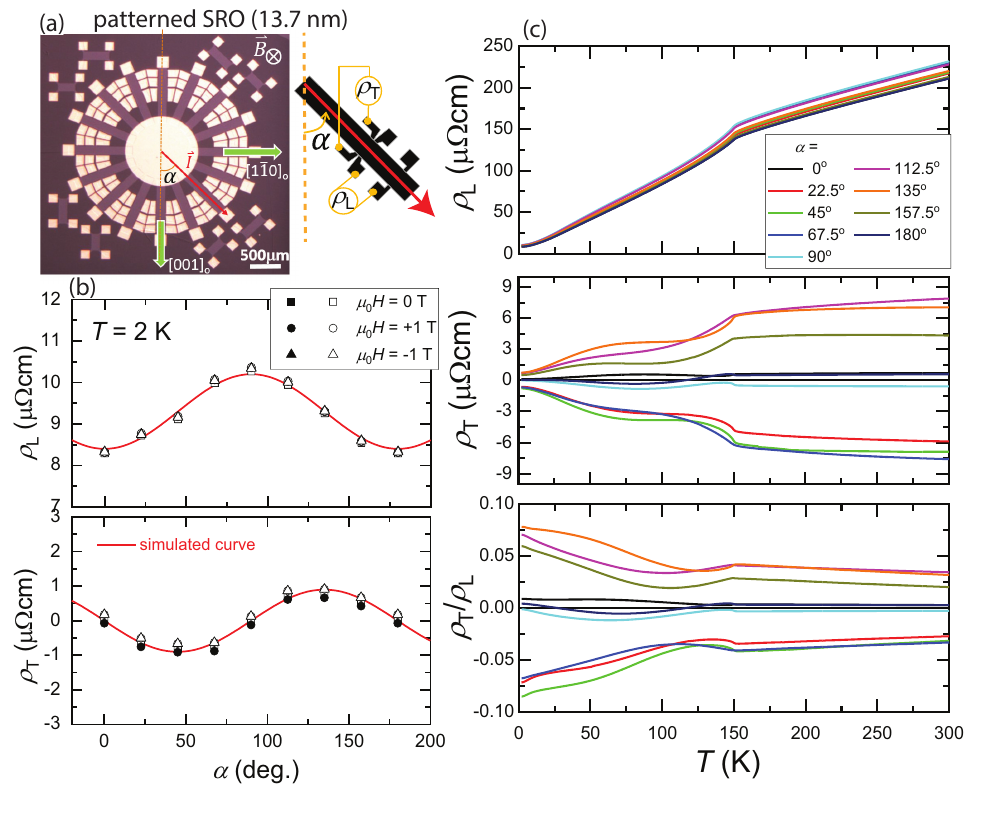}
  \caption{ The resistivity anisotropy in an untwinned SRO thin film. (a) An optical image of a patterned sunbeam SRO device with a film thickness of $t \approx$ 13.7 nm. The white scale bar is 500 $\mu$m. Green arrows indicate the SRO orthorhombic crystalline directions. The right panel illustrates a Hall bar device with the definitions for the measured longitudinal and transverse resistivity as $\rho_{\rm L}$ and $\rho_{\rm T}$, respectively. $\alpha$ angle is defined as the angle between the bias current direction (red arrow) and [001]$_{\rm o}$. (b) The $\alpha$-dependent $\rho_{\rm L}$ and $\rho_{\rm T}$ for three different field values of 0, -1, and +1 T. The experimental data agree well with the simulated curves (red curves) based on a resistivity anisotropy model. The upper, middle and lower panels of (c) show the $T$-dependent $\rho_{\rm L}$, $\rho_{\rm T}$, and the $\rho_{\rm T}/\rho_{\rm L}$ ratio, respectively, for different Hall bar devices with $\alpha$ values ranging from 0$^{\rm o}$ to 180$^{\rm o}$. See the text for more detailed descriptions.     
  }
  \label{device}
\end{figure}

Figure \ref{device}(a) shows an optical image of a sunbeam device patterned on an untwinned SRO thin film with a thickness $t$ of about 13.7 nm. By using a STO (001) substrate with a miscut angle of about 0.1 degrees along one of the principal cubic axes, the volume fraction of the dominant domain was determined by high resolution X-ray scattering via the (02$\pm$1)$_{\rm o}$ reflections to be about 95 \% \cite{Kar2021} (see Supplementary Note 1 \cite{SOM}), where the orthorhombic crystalline directions are shown in Fig. \ref{device}(a). The right panel of Fig. \ref{device}(a) illustrates one of the Hall bars in the sunbeam device, and $\alpha$ defines the angle between the bias current direction and [001]$_{\rm o}$. $\rho_{\rm L}$ and $\rho_{\rm T}$ correspond to the longitudinal and transverse resistivity, respectively. With a compressive strain of about -0.4 \%, the Curie temperature $T_{\rm c}$ for SRO thin film is about 150 K, the magnetic easy axis is close to the film surface normal of [110]$_{\rm o}$ \cite{Koster2012,Kar2021}. From rigorous angle dependent coercive field measurements on SRO thin film with $t$ = 13.7 nm at low temperature (see Supplementary Note 2 \cite{SOM}), we identified the magnetic easy axis to be about 20$^{\rm o}$ away from the [110]$_{\rm o}$ and tilting toward [\=110]$_{\rm o}$. Fig. \ref{device}(b) shows the $\alpha$-dependent $\rho_{\rm L}$ and $\rho_{\rm T}$ values at three different applied field values of 0, -1, and +1 T along [110]$_{\rm o}$ at $T$ = 2 K. The solid (open) symbols in Fig. \ref{device}(b) correspond to the data extracted from field sweeping of +5 to -5T (-5 to +5 T). The $\rho_{\rm L}$ appears to be at a maximum value of about 10.4 $\mu\Omega$cm for $\alpha$ = 90$^{\rm o}$ and drops to a minimum value of about 8.1 $\mu\Omega$cm for $\alpha$ = 0 and 180$^{\rm o}$, exhibiting a clear cos(2$\alpha$) dependence. On the other hand, the  $\rho_{\rm T}$ shows a sin(2$\alpha$) dependence instead with a maximum magnitude of about 0.9 $\mu\Omega$cm at $\alpha$ = 45$^{\rm o}$ and 135$^{\rm o}$. The simulated curves using a resistivity anisotropy model of $\rho_{\rm L}$ and  $\rho_{\rm T}$ are shown as red curves in Fig. \ref{device}(b) (see Supplementary Note 3 \cite{SOM}). We note that the amplitude of the anisotropy is significantly larger than the small changes in $\rho_{\rm L}$ and $\rho_{\rm T}$ when reversing the magnetization by changing the field from +1 to -1 T, inferring that the observed resistivity anisotropy in our SRO thin films is not dictated by the magnetic domain wall resistance and other magnetization-related effects. The upper, middle, and lower panels of Fig. \ref{device}(c) show the temperature dependence of $\rho_{\rm L}$, $\rho_{\rm T}$, and $\rho_{\rm T}/\rho_{\rm L}$ ratio, respectively, for different $\alpha$ values ranging from 0$^{\rm o}$ to 180$^{\rm o}$. The residual resistivity ratio of $\rho_{\rm L}(300K)/\rho_{\rm L}(5K)$ varies weakly and equals about 24.0 and 21.4 for $\alpha$ = 0$^{\rm o}$ and 90$^{\rm o}$, respectively. Those results support the nearly single structure domain and thus untwinned nature in our SRO thin films, and also the exact dimensions of each Hall bar at different $\alpha$ values are very close to each other, which justifies the feasibility for the investigation of anisotropy effects in our SRO thin films. As the $T$ decreases, we note that the magnitude of $\rho_{\rm T}/\rho_{\rm L}$ ratio for $\alpha$ = 45$^{\rm o}$ slightly decreases near the  $T_{\rm c}$ and then increases again below 100K, attaining a sizable ratio of $\rho_{\rm T}/\rho_{\rm L} \approx$ -0.085 at $T$ = 2 K without saturation.     

\begin{figure}
\includegraphics[width=\linewidth]{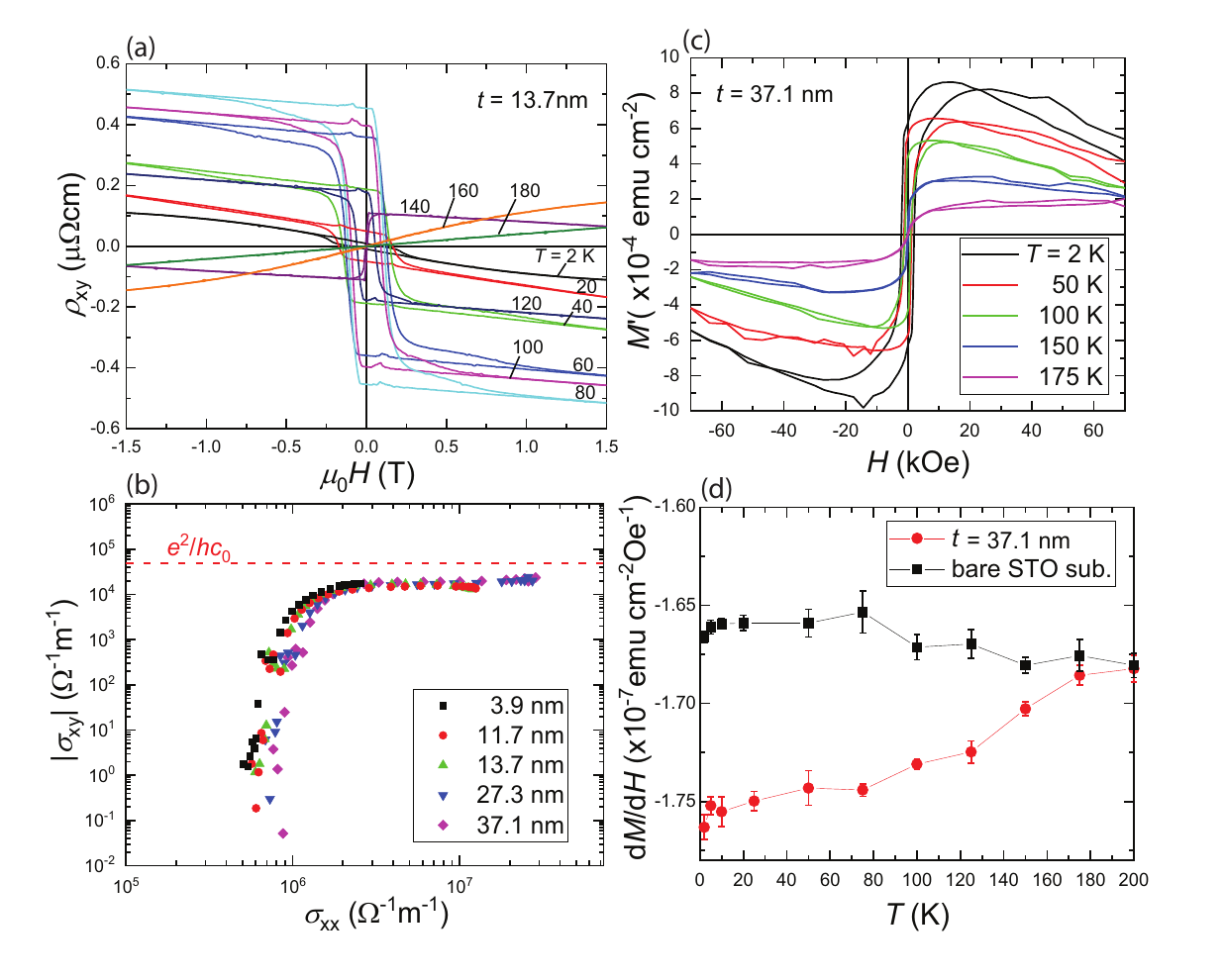}
  \caption{Intrinsic AHE and enhanced diamagnetism in SRO thin films. (a) The hysteresis loops of Hall resistivity $\rho_{\rm xy}$ at different temperatures ranging from 2 to 180 K for a SRO thin film with $t \approx$ 13.7 nm. (b) The magnitude of the Hall conductivity at zero field $|\sigma_{\rm xy}|$ is plotted as a function of the corresponding zero-field conductivity $\sigma_{\rm xx}$ for SRO thin films with different $t$s ranging from 3.9 nm to 37.1 nm. $|\sigma_{\rm xy}|$ at low temperatures approach a constant value of about 2.0 $\times 10^{4} \Omega^{-1}m^{-1}$, which is relatively close to the intrinsic anomalous Hall conductivity of $e^2/hc_{0} \approx$ 5.0 $\times 10^{4} \Omega^{-1}m^{-1}$. (c) The field dependence of background subtracted magnetization $M'$ for a thicker SRO film with $t \approx$ 37.1 nm, exhibiting an increased diamagnetic response for $H \geq$ 2T as $T$ drops. (d) The $T$ dependence of averaged slope of $dM/dH$ (red circles) in high field regime from 2 to 7 T shows a progressive decrease with decreasing $T$, indicating an enhanced diamagnetic response in SRO film at low $T$, which is in big contrast to the nearly $T$-independent slope of $dM/dH$ (black square) for a bare STO substrate.      
  }
  \label{AHE}
\end{figure}

Now, we turn to the discussions about the anomalous Hall effect (AHE) and the magnetization data in our SRO thin films. Figure \ref{AHE}(a) shows the field dependent Hall resistivity $\rho_{\rm xy}$ at different temperatures ranging from 2 to 180 K, where weak field hysteresis loops in $\rho_{\rm xy}$-$\mu_{\rm 0}H$ curves with a small coercive field of less than 0.2 T were observed below $T_{\rm c}$ as expected. The magnitude of converted Hall conductivity $|\sigma_{\rm xy}|$ at zero field is plotted in Fig. \ref{AHE}(b) as a function of the corresponding conductivity $\sigma_{\rm xx}$ in logarithmic scales for SRO thin films with different thicknesses $t$s ranging from 3.9 to 37.1 nm. Remarkably, $|\sigma_{\rm xy}|$ appears to approach a constant and $t$-independent value of about 2.0 $\times 10^{4}$ $\Omega^{-1}m^{-1}$ at low temperatures, which falls in the same order as the intrinsic anomalous Hall conductivity due to the Berry curvatures of the bulk band, i.e., $e^2/hc_{\rm o} \approx$ 5.0 $\times 10^{4}$ $\Omega^{-1}m^{-1}$ ($c_{\rm o}$ being the orthorhombic lattice constant of about 7.81\AA) shown as the red dashed line in Fig. \ref{AHE}(b). We note insignificant changes in the $|\sigma_{\rm xy}|$ with $\sigma_{\rm xx}$ down to $T$ = 1.4 K, and this thus suggests a negligible contribution from the extrinsic skew scattering effect to AHE, where a linear relation of $|\sigma_{\rm xy}| \propto \sigma_{\rm xx}$ is expected instead \cite{Nagaosa2010}. On the other hand, rigorous magnetization measurements were performed on a thicker SRO film with $t \approx$ 37.1 nm using a superconducting quantum interference device (SQUID) magnetometer. By subtracting the diamagnetic background at 200 K, the resulting magnetization $M' - H$ curves at different temperatures are shown in Fig. \ref{AHE}(c), where, for $\mu_{0}H \geq$ 2 T, the diamagnetic response seems to increase as the temperature drops. As shown in Fig. \ref{AHE}(d), the averaged slope of $dM/dH$ for the field regime from $\mu_{0}H \geq$ 2 T to 7 T is negative with increasing magnitude as the temperature decreases to 2 K, which is in big contrast to the nearly $T$-independent slope from the controlled measurements on a bare STO substrate (square symbols in Fig. \ref{AHE}(d)). The observed intrinsic $|\sigma_{\rm xy}| \sim e^2/hc_{\rm o}$ \cite{Fang2003,Chen2013} and the enhanced diamagnetic response \cite{Rao2014,Sue2021} at low temperatures strongly support the presence of the Weyl-nodes near the Fermi surface and thus the Weyl metal phase in SRO. We also remark that the zero-field Hall signals at low temperatures in SRO are dominated by the intrinsic AHE, which would be important for the subsequent discussions about the observed nonlinear Hall signals in SRO. 

\begin{figure}
\includegraphics[width=\linewidth]{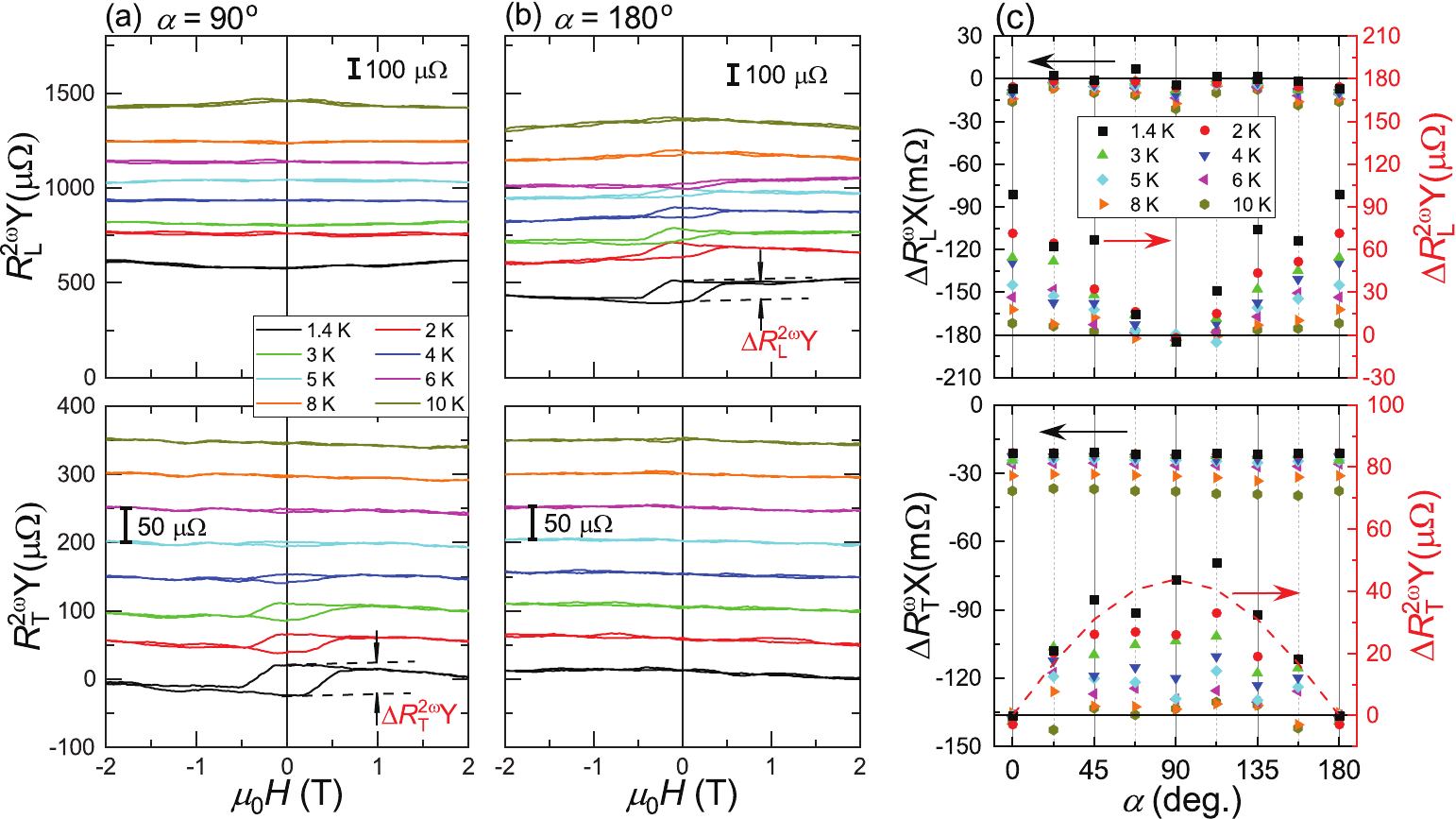}
  \caption{The angular dependent nonlinear and nonreciprocal signals in SRO sunbeam device with $t \approx$ 13.7 nm. The upper and lower panels of (a) show the $R_{\rm L}^{2\omega}$ and $R_{\rm T}^{2\omega}$, respectively, at eight different $T$s ranging from 1.4 K to 10 K for $\alpha$ = 90$^{\rm o}$ Hall bar device. For $T\geq$ 2 K, the curves of $R_{\rm L}^{2\omega} - \mu_{\rm 0}H$ and $R_{\rm T}^{2\omega} - \mu_{\rm 0}H$ are systematically shifted upward by multiple of 100 and 50 $\mu\Omega$, respectively, for clarity. Similarly, (b) shows the results for $\alpha$ = 180$^{\rm o}$ Hall bar device, sharing the same color codes and also axes labels. The $\Delta R_{\rm T}^{2\omega}$ and $\Delta R_{\rm L}^{2\omega}$ at zero magnetic field are defined in the lower and upper panel of (a) and (b), respectively. A summary of the $\alpha$ dependent $\Delta R_{\rm L}^{\omega}$ and $\Delta R_{\rm L}^{2\omega}$ ($\Delta R_{\rm T}^{\omega}$ and $\Delta R_{\rm T}^{2\omega}$) are shown in the upper (lower) panel of (c). The resulting $\Delta R_{\rm T}^{2\omega}$ data fall relatively close to a simple sin$\alpha$ curve shown as the red dashed line.       
  }
  \label{NLH}
\end{figure}

As illustrated in the right panel Fig. \ref{device}(a), the second harmonic longitudinal ($R_{\rm L}^{2\omega}$) and transverse ($R_{\rm T}^{2\omega}$) resistance were measured with a bias current of 0.7 mA at a frequency of about 18.4 Hz. The resulting complex second harmonic signal can be expressed as $\tilde{R}_{\rm L(T)}^{2\omega}$ = $R_{\rm L(T)}^{2\omega}$X + $i$ $R_{\rm L(T)}^{2\omega}$Y, which is probed by a lock-in amplifier. The upper and lower panel of Fig. \ref{NLH}(a) shows the field dependent $R_{\rm L}^{2\omega}$Y and $R_{\rm T}^{2\omega}$Y, respectively, for $\alpha$ = 90$^{\rm o}$ Hall bar device at different $T$s ranging from 1.4 K to 10 K. For clarity, the curves of $R_{\rm L}^{2\omega}$Y$ - \mu_{0}H$ and $R_{\rm T}^{2\omega}$Y$ - \mu_{0}H$ at different $T$s are systematically shifted upward by multiple of 100 $\mu\Omega$ and 50 $\mu\Omega$, respectively. For $T \geq$ 10K, both $R_{\rm L}^{2\omega}$Y and $R_{\rm T}^{2\omega}$Y show no hysteresis loops in the weak field regime, which is in big contrast to the sizable $\rho_{\rm xy} - \mu_{0}H$ loops shown in Fig. \ref{AHE}(a) at similar temperatures. Below 6K, a sizable hysteresis loop starts to appear in $R_{\rm T}^{2\omega}$Y as shown in the lower panel of Fig. \ref{NLH}(a), but $R_{\rm L}^{2\omega}$Y remains nearly field-independent without showing a hysteresis loop. The definition of $\Delta R_{\rm T}^{2\omega}$Y is illustrated in the lower panel of Fig. \ref{NLH}(a), and it corresponds to the change of the $R_{\rm T}^{2\omega}$Y signal at zero magnetic field when reversing the magnetization of the SRO thin film. For $\alpha$ = 90$^{\rm o}$ Hall bar device with bias current $I$ along [1\=10]$_{\rm o}$, the $\Delta R_{\rm T}^{2\omega}$Y gradually increases in magnitude as $T$ drops, giving a $\Delta R_{\rm T}^{2\omega}$Y $\approx$ 44 $\mu\Omega$ at $T =$ 1.4 K. Remarkably, for $\alpha$ = 180$^{\rm o}$ Hall bar device with a bias current $I$ along [001]$_{\rm o}$ as demonstrated in Fig. \ref{NLH}(b), the hysteresis loops appear in the longitudinal channel of $R_{\rm L}^{2\omega}$Y at low temperatures instead, giving a value of $\Delta R_{\rm L}^{2\omega}$Y $\approx$ 100 $\mu\Omega$ at $T =$ 1.4 K, and no hysteresis loops were observed in the transverse channel ($R_{\rm T}^{2\omega}$Y). 

Figure \ref{NLH}(c) summarized the results from 9 different $\alpha$ values Hall bars from the sunbeam device shown in Fig. \ref{device}(a) (see Supplementary Note 4 for detailed descriptions on measurement geometry and polarity \cite{SOM}). The upper panel of Fig. \ref{NLH}(c) shows the first harmonic signals ($\Delta R_{\rm L}^{\omega}$X) and second harmonic signals ($\Delta R_{\rm L}^{2\omega}$Y) in the longitudinal channel as a function of $\alpha$ with different $T$s. $\Delta R_{\rm L}^{2\omega}$Y exhibits a maximum value of about 100 $\mu\Omega$ at $\alpha$ = 0$^{\rm o}$ and 180$^{\rm o}$, and it gradually decreases in magnitude to zero as $\alpha$ approaches 90$^{\rm o}$. In contrast, the first harmonic signals of $\Delta R_{\rm L}^{\omega}$X are nearly zero for all $\alpha$ and $T$ values as expected. On the other hand, the lower panel of Fig. \ref{NLH}(c) puts together the $\alpha$ dependent first harmonic signals ($\Delta R_{\rm T}^{\omega}$X) and second harmonic signals ($\Delta R_{\rm T}^{2\omega}$Y) in the transverse channel at different $T$s. Unlike the longitudinal channel, the $\Delta R_{\rm T}^{2\omega}$Y data show a relatively good agreement to the sin$\alpha$ dependence [red dashed line in the lower panel of Fig. \ref{NLH}(c)], giving a value of $\Delta R_{\rm T}^{2\omega}$Y $\approx$ 44 $\mu\Omega$ at $\alpha$ = 90$^{\rm o}$ and vanishing values for $\alpha$ = 0$^{\rm o}$ and 180$^{\rm o}$. Such a unique sin$\alpha$ dependence in $\Delta R_{\rm T}^{2\omega}$Y is drastically distinct from the nearly $\alpha$-independent first harmonic signals of $\Delta R_{\rm T}^{\omega}$X.                  

\begin{figure}
\includegraphics[width=\linewidth]{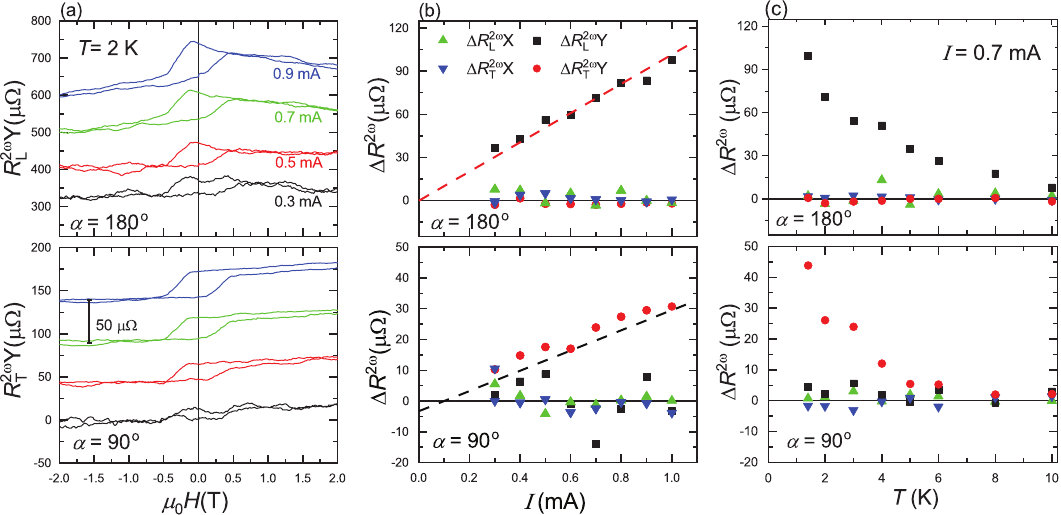}
  \caption{ The $I$ and $T$ dependences of second harmonic signals for SRO sunbeam device with $t \approx$ 13.7 nm. Upper panel of (a) shows the $R_{\rm L}^{2\omega}$Y $- \mu_0H$ curves for $\alpha$ = 180$^{\rm o}$ with four different $I$s ranging from 0.3 mA to 0.9 mA. Similarly, the lower panel of (a) shows the $R_{\rm T}^{2\omega}$Y $- \mu_0H$ curves for $\alpha$ = 90$^{\rm o}$ with four different $I$s. Two panels share the same color codes, and the curves are shifted upward for clarity. The corresponding $I$-dependent second harmonic signals of $\Delta R_{\rm L(T)}^{2\omega}$X and $\Delta R_{\rm L(T)}^{2\omega}$Y for $\alpha$ = 180$^{\rm o}$ and 90$^{\rm o}$ are shown in upper-panel and lower-panel, respectively, of (b). For $\alpha$ = 180$^{\rm o}$, only $\Delta R_{\rm L}^{2\omega}$Y is sizable with practical $I$-linear dependence. On the contrary, for $\alpha$ = 90$^{\rm o}$ with bias current along [1\=10]$_{\rm o}$, $\Delta R_{\rm L}^{2\omega}$Y becomes nearly zero with no apparent $I$ dependence, but instead the transverse channel of $\Delta R_{\rm T}^{2\omega}$Y bears a clear $I$-dependence, inferring the presence of an effective $\vec D$ along [1\=10]$_{\rm o}$.  (c) shows the progressive increases in the magnitude with decreasing $T$ for both $\Delta R_{\rm L}^{2\omega}$Y of $\alpha$ = 180$^{\rm o}$ and $\Delta R_{\rm T}^{2\omega}$Y of $\alpha$ = 90$^{\rm o}$. (b) and (c) share the same symbols and color codes.      
  }
  \label{NLHIdep}
\end{figure}

For consistency check, the current dependent $R_{\rm L}^{2\omega}$Y for $\alpha$ = 180$^{\rm o}$ and $R_{\rm T}^{2\omega}$Y for $\alpha$ = 90$^{\rm o}$ at $T$ = 2 K are shown in the upper panel and lower panel, respectively, of Fig. \ref{NLHIdep}(a) with different bias currents ranging from 0.3 to 0.9 mA, where the curves are systematically shifted upward for clarity. For $\alpha$ = 180$^{\rm o}$, $\Delta R_{\rm L}^{2\omega}$Y progressive increases from 35 to 84 $\mu\Omega$ as the bias current $I$ increases from 0.3 to 0.9 mA. The detailed $I$-dependent second harmonic signals ($\Delta R_{\rm L(T)}^{2\omega}$X + $i$ $\Delta R_{\rm L(T)}^{2\omega}$Y) were shown in the upper panel of Fig. \ref{NLHIdep}(b), where only $\Delta R_{\rm L}^{2\omega}$Y data show nearly $I$-linear dependent behavior, and all other second harmonic signals are vanishingly small. On the contrary, for $\alpha$ = 90$^{\rm o}$, the $\Delta R_{\rm T}^{2\omega}$Y increases from about 10 to 30 $\mu\Omega$ as $I$ increases from 0.3 to 0.9 mA, and the corresponding $I$-dependent signals are shown in the lower panel of Fig. \ref{NLHIdep}(b). The nearly $I$-linear dependence of $\Delta R_{\rm T}^{2\omega}$Y for $\alpha$ = 90$^{\rm o}$ only appears in the transverse channel but not in the longitudinal channel of $\Delta R_{\rm L}^{2\omega}$Y, justifying the presence of nonlinear Hall effect in SRO thin films. The magnitude of both $\Delta R_{\rm L}^{2\omega}$Y for $\alpha$ = 180$^{\rm o}$ and $\Delta R_{\rm T}^{2\omega}$Y for $\alpha$ = 90$^{\rm o}$ grow rapidly as $T$ drops below 10 K as shown in the upper panel and lower panel, respectively, of Fig. \ref{NLHIdep}(c), which is dramatically different from the minor drops in $\rho_{L(T)}$ and the nearly constant $\sigma_{\rm xy}\equiv \rho_{T}/(\rho_{L}^2+\rho_{T}^2$) with decreasing $T$ as shown in Fig. \ref{device}(c) and Fig. \ref{AHE}(b), respectively. We also note that the extracted $\Delta R_{\rm T}^{2\omega}$ and $\Delta R_{\rm L}^{2\omega}$ do not vary significantly with the bias current frequency (see Supplementary Note 5 \cite{SOM}), and they derive from the difference in the second harmonic signals between opposite magnetization directions in SRO at zero external magnetic field as illustrated in Figs. \ref{NLH}(a) and (b). Therefore, the extrinsic contact effects and also possible magnetic field related effects for NRTE and nonlinear Hall effects can be excluded \cite{Morimoto2016,LiRH2021,Nandy2021}.                
             
\section{Discussions}

In a magnetic system with broken time reversal symmetry, both intrinsic and extrinsic AHE can contribute to the measured Hall signals \cite{Nagaosa2010}, and nonlinear Hall signals at the second harmonic generally require additional inversion symmetry breaking \cite{Gao2014,Sodemann2015,Du2019,Iso2020,He2021}. As demonstrated in Fig. \ref{AHE}(b), the low-temperature AHE in SRO was dominated by the contribution from the intrinsic AHE due to Weyl nodes near the Fermi-surface \cite{Fang2003}, where $\sigma_{\rm xy}$ is nearly a constant of about $e^2/hc_{\rm o}$ down to about 1.4 K, and thus extrinsic skew scattering effect \cite{Iso2020,He2021} shall not play a significant role for our observed nonlinear Hall signals. On the other hand, the distinct sin$\alpha$ dependence of $\Delta R_{\rm T}^{2\omega}$ does not seem to be compatible with the intrinsic mechanism due to the electron-lifetime-independent quantum metric dipole (QMD) \cite{Gao2014,Wang2021,Liu2021,Gao2023,Wang2023}, where intrinsic AHE at zero field ($\Delta R_{\rm T}^{\omega}$) is nearly $\alpha$ independent as shown in the lower panel of Fig. \ref{NLH}(c). Therefore, the observed nonlinear Hall signals of $\Delta R_{\rm T}^{2\omega}$Y is more likely derived from the BCD \cite{Sodemann2015} due to bulk or surface states with inversion symmetry breaking.

The BCD contribution to the second harmonic current density can be derived as $j_{a}^{2\omega} =  \chi_{abc} E_{b}E_{c}$, and $\chi_{abc} \equiv -\varepsilon_{adc} \frac{e^3\tau}{2\hbar^2(1+i\omega\tau)}D_{bd}$. The BCD can be expressed as $D_{bd} \equiv \int \frac{d^3k}{(2\pi)^3}f_{0}\frac{\partial \Omega_{d}}{\partial k_{b}}$, where $f_0$ and $\Omega$ are the equilibrium Fermi-Dirac distribution and the Berry curvature, respectively, and it can be nonzero for systems with titled Weyl nodes and inversion asymmetry \cite{Sodemann2015,Zhang2018,Du2018}. Therefore, with a bias current along $b$ axis, the resulting nonlinear Hall current is simply $j_{a}^{2\omega} = \chi_{abb} E_{b}^2$ with $\chi_{abb} =  \frac{e^3\tau}{2\hbar^2(1+i\omega\tau)}D_{bc}$, and thus $j_{a}^{2\omega}$ is a direct measure for the Berry curvature gradient along the bias current direction. In our sunbeam device with different bias current directions of $\alpha$ values ranging from 0$^{\rm o}$ to 180$^{\rm o}$, the largest nonlinear Hall signal was observed with $\alpha = 90^{\rm o}$, inferring the presence of an effective BCD $\vec {D}$ along [1\=10]$_{\rm o}$. In order to compare the magnitude of our observed nonlinear Hall effect with other systems, we adopted the 3D formula with resistivity anisotropy effect shown in Fig. \ref{device}(b). The $\alpha$ dependent $\Delta R_{\rm T}^{2\omega}$ can be deduced to give $\Delta R_{\rm T}^{2\omega} = \frac{\chi_{abb}\rho_{a}\rho_{b}^2}{Wt^2} I$sin$\alpha$, where $\rho_{b}$($\rho_{a}$) is the resistivity along [1\=10]$_{\rm o}$([001]$_{\rm o}$), and $W$ is the width of the Hall bar device ($W$ = 150 $\mu$m) (see Supplementary Note 2 \cite{SOM}). The sin$\alpha$ and $I$-linear dependences of $\Delta R_{\rm T}^{2\omega}$ are well confirmed by the experiment shown in lower-panel of Fig. \ref{NLH}(c) and Fig. \ref{NLHIdep}(b), respectively. By using a Drude electron lifetime of about $\tau_d \sim$ 1.9 $\times$ 10$^{-13}$ s, the magnitude of the effective 3D BCD can be roughly estimated to be about $|\vec {D}| \approx $ 55, which falls in the same order of magnitude as several other reported 3D Weyl systems with large BCD \cite{Zhang2022}, and the corresponding $E_{\rm y}^{2\omega}$/($E_{\rm x}^{\omega})^2 \approx $ 1.7 $\times$ 10$^{-7}$ (m/V) at $T =$ 2 K. 

\begin{figure}
\includegraphics[width=12cm]{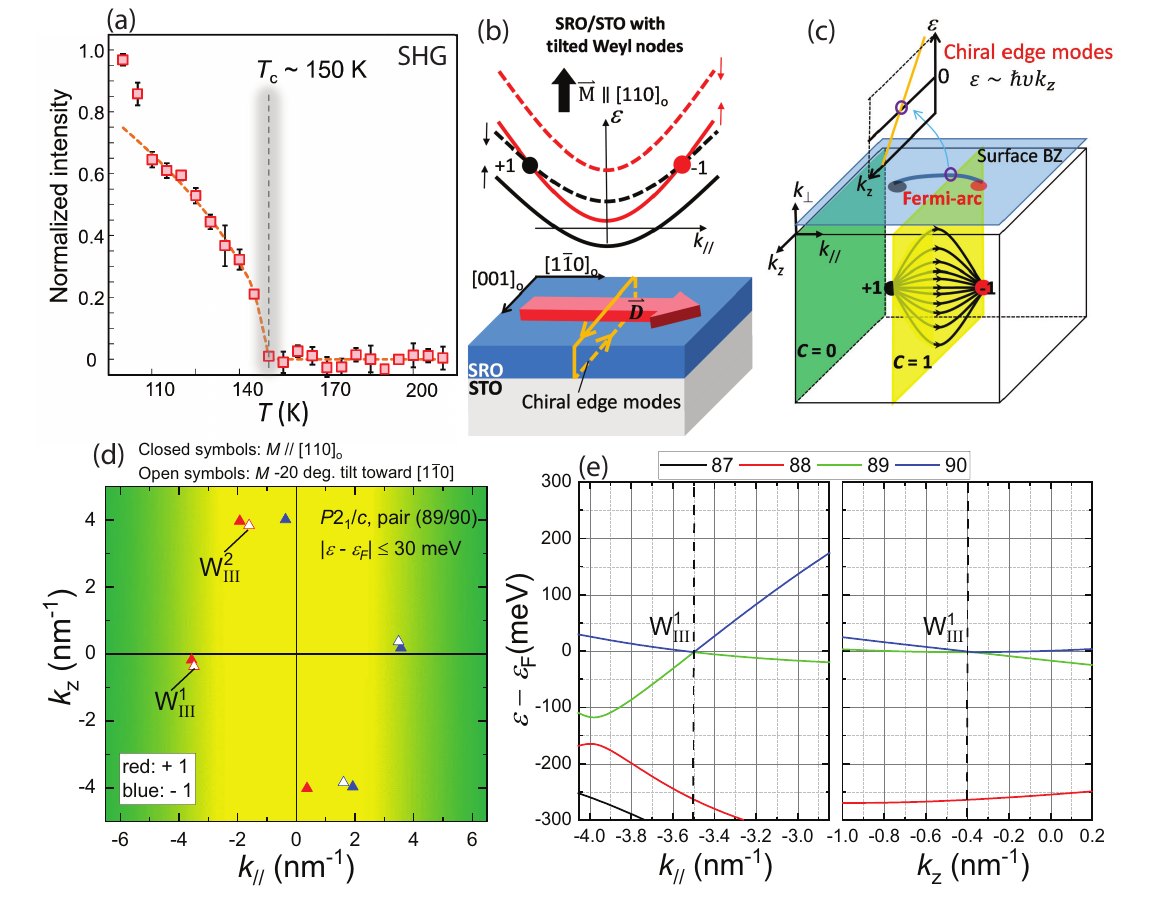}
  \caption{ A proposed scenario for the observed NRTE in Weyl metal thin film of SRO/STO. (a) The temperature dependence of the scattering plane-averaged SHG intensity from a SRO/STO film with $t$ $\approx$ 35 nm measured in $S_{\rm in}$ – $S_{\rm out}$ polarization geometry. The dashed grey line marks the Curie temperature $T_{\rm c}$. A constant background based on the SHG intensity above $T_{\rm c}$ \cite{Roh2021} was subtracted off. The orange curve is a guide to the eye. The error bar shows the standard deviation of the intensities over 5 independent measurements. (b) shows the proposed real space scenario of the SRO/STO system with tilted Weyl nodes, giving rise to an effective BCD $\vec{D}$ from surface states along [1\=10]$_{\rm o}$ and 1D chiral edge modes along [001]$_{\rm o}$. (c) An illustration of a minimum model of Weyl system with one Weyl-node pair of chiral charges +1 and -1. For 2D slices between the Weyl-node pair, it gives a Chern number 1 with 1D chiral edge modes on the boundary of the system as demonstrated in the upper panel of (c). (d) Calculated Weyl nodes locations projected on $k_{//}$ -$k_{\rm z}$ plane, and only Weyl nodes with energies of $|\varepsilon-\varepsilon_{\rm F}| \leq$ 30 meV are included. Open and closed symbols correspond to different magnetization orientation condition, and the red (blue) color indicates the corresponding chiral charge of +1 (-1). The shaded yellow region indicates the non-zero total Chern number, supporting the existence of 1D chiral edge modes along $k_{\rm z}$. (e) The band dispersions along $k_{//}$ and $k_{\rm z}$ for Weyl-node of W$_{\rm III}^1$ show a tilted Weyl-node.
}
\label{cal}
\end{figure}

For bulk inversion symmetry consideration in SRO thin films, the onset of ferromagnetism for $T \leq$ 150 K with magnetization along [110]$_{\rm o}$ can, in principle, break the mirror planes with normal vectors perpendicular to the magnetization direction, and a similar mirror symmetry breaking by magnetism has been reported before \cite{Torre2021}. The absence of the mirror plane at low temperature was further confirmed by the change of the crystal structure to monoclinic phase with a space group of $P2_1/c$ for $T \leq$ 150 K, which was determined by high precision X-ray measurements down to 10 K (see Supplementary Note 1 \cite{SOM}). We also conducted rotational anisotropy second harmonic generation (SHG) measurements, which can be sensitive to the magnetic order parameter in perovskite transition metal oxides \cite{Seyler2020}. Figure \ref{cal}(a) shows the temperature dependence of the scattering plane angle averaged SHG intensity from a SRO/STO film with $t \approx$ 35 nm, which exhibits an intensity upturn below 150 K. Although we did not resolve whether the magnetic order induced SHG susceptibility is directly proportional to the magnetization or to its square (as would be the case for magnetostriction), the critical temperature is consistent with that reported for bulk single crystals. We also noted a progressive increase in the SHG intensity as temperature decreases further. We remark that the low temperature space group $P2_1/c$ lacks mirror planes but remains inversion symmetric, which excludes significant bulk structure inversion symmetry breaking due to possible lattice strain gradient effect \cite{Hwang2012,Pesq2012,Sohn2021}. Nevertheless, due to strong $d$-$p$ hybridization, the calculated O moments turn out to contribute up to about $20 \%$ of the total magnetic moment in SRO, and thus the resulting non-collinear magnetic configurations \cite{mSHG1,mSHG2,Taki2020} (see also Supplementary Note 6) can, in principle, give rise to a magnetic space group with a broken bulk inversion symmetry and thus non-zero bulk BCD. From rigorously calculated band dispersions along $k_{\rm //}$ and $k_{\rm z}$ (see Supplementary Note 6 \cite{SOM}), we found that all Weyl nodes appear to be tilted and thus can give rise to nonzero BCD with broken bulk inversion symmetry \cite{Zhang2018}. Taking Weyl node of $W_{\rm III}^1$ with $|\varepsilon-\varepsilon_{\rm F}|$= 1.87 meV as an example, the band dispersions along $k_{//}$ and $k_{\rm z}$ are plotted in the left panel and right panel, respectively, of Fig. \ref{cal}(e). It shows a large tilting of Weyl node along both $k_{\rm //}$ and $k_{\rm z}$. Nevertheless, our calculated bulk BCD based on such a non-collinear magnetic configuration only gives a $|\vec {D}| \leq $ 10$^{-2}$ that is too small by orders of magnitude to explain our observed effective $|\vec {D}|$ (see Supplementary Note 7 \cite{SOM}). We also performed the bulk QMD calculations, and the resulting nonlinear QMD conductivity ($\sigma_{abc}$ = $j_{a}^{2\omega}/(E_{b}^{\omega}E_{c}^{\omega}$)) has a magnitude of $\leq$ 30 $\mu$A/V$^2$ that corresponds to a $E_{\rm y}^{2\omega}$/($E_{\rm x}^{\omega})^2 \leq $ 10$^{-11}$ (m/V), which is more than four orders of magnitude smaller than our observed nonlinear Hall signals (see Supplementary Note 7). While the tilted bulk Weyl nodes in SRO with a non-collinear magnetic configuration can give rise to non-zero bulk BCD and QMD contributions to nonlinear Hall signals, their calculated magnitudes are too small to account for the observed $\Delta R_{\rm T}^{2\omega}$ in our experiment.            

\begin{figure}
\includegraphics[width=\linewidth]{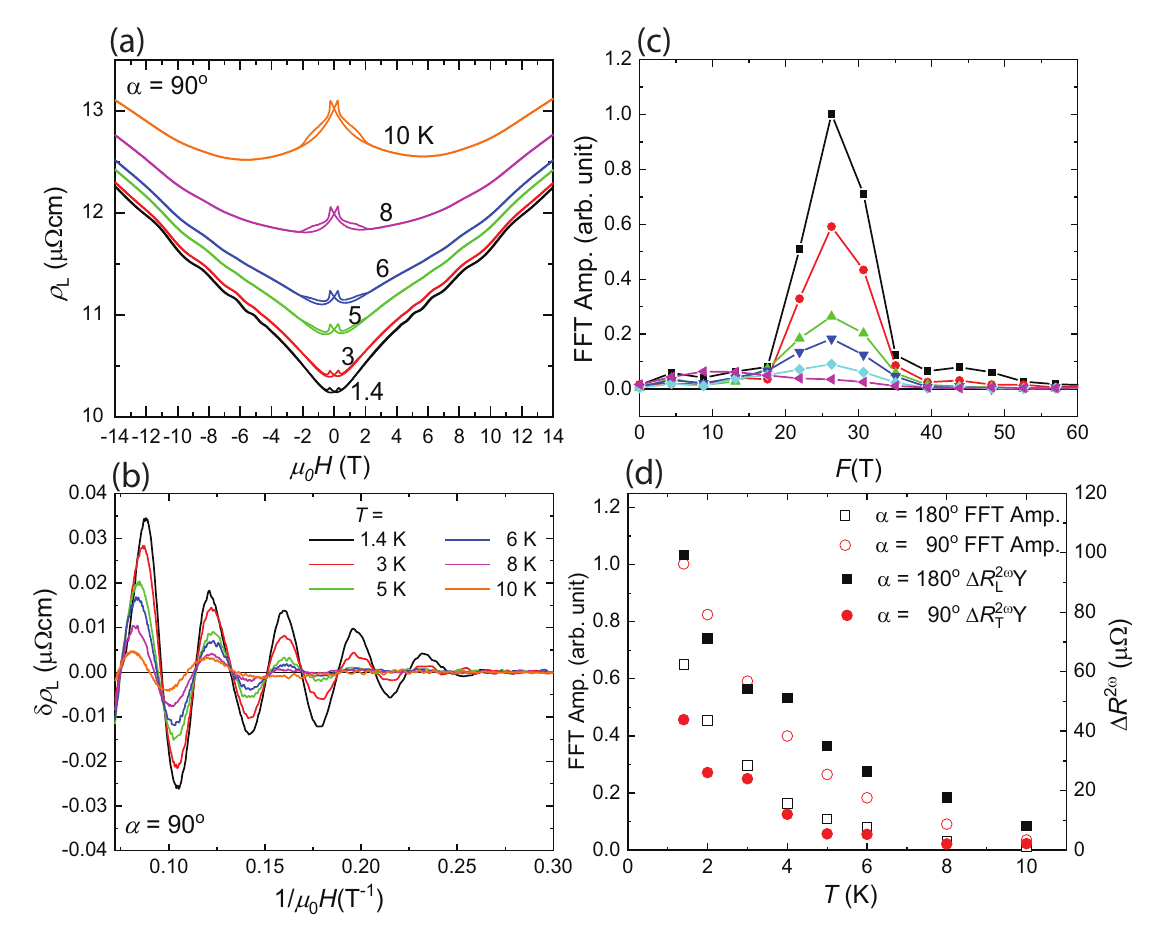}
  \caption{The transverse magnetoresistance crossover and the emergence of quantum oscillations below 10 K. (a) A crossover of transverse MR from negative to positive in the weak field regime as $T$ decreases from 10 to 1.4 K. (b) The $T$ dependence of pure oscillating component of $\rho_{\rm L}$ ($\delta \rho_{\rm L}$) as a function of $1/\mu_{0}H$ for $\alpha$ = 90$^{\rm o}$, and the corresponding FFT spectra are shown in (c) that shares the same color code as (b). It reveals a dominant contribution from a 2D-like quantum oscillation with a frequency of about 28 T, where its oscillating amplitude grows rapidly below 10 K. The $T$ dependence of FFT amplitudes, $\Delta R_{\rm T}^{2\omega}$, and $\Delta R_{\rm L}^{2\omega}$ for $\alpha$ = 90$^{\rm o}$ and 180$^{\rm o}$ are shown in (d), indicating a close connection of the observed $\Delta R_{\rm L(T)}^{2\omega}$ to the transverse MR crossover and 2D-like quantum oscillations below 10 K, where the surface dominant charge transport plays an important role.          
  }
  \label{surface}
\end{figure}

On the other hand, surface projected Weyl nodes and also Fermi-arc surface states, lacking inversion symmetry, may also give rise to nonzero BCD $\vec{D}$. The growing contribution from surface states at low temperatures is in accord with the dramatic changes of magnetotransport behavior below 10 K as demonstrated in Fig. \ref{surface}. As $T$ decreases from 10 K to 1.4 K, the weak field transverse MR with field along the [110]$_{\rm o}$ shows a crossover from a negative transverse MR to a positive transverse MR as shown in Fig. \ref{surface}(a), and the Hall resistivity (Fig. \ref{AHE}(a)) also shows a nonlinear field dependence below 10 K, indicating a multiple channel conduction at lower temperatures. On the other hand, pronounced quantum oscillations with a frequency of about 28 T were observed for all $\alpha$ values in our sunbeam device as shown in Fig. \ref{surface} (b) for $\alpha$ = 90$^{\rm o}$, and the corresponding Fast Fourier transform (FFT) spectra for different $T$s were shown in Fig. \ref{surface}(c). We note that 28 T quantum oscillations in SRO thin film were recently reported to behave as a 2D-like Fermi pocket with signatures that are consistent with Weyl-orbit quantum oscillation effect due to the bulk tunneling between the top and bottom Fermi-arc surface states \cite{Kaneta2022,kar2022}. The open black squares and open red circles in Fig. \ref{surface}(d) plot the rapid increase of FFT amplitude for quantum oscillations below 10 K for $\alpha$ = 180$^{\rm o}$ and 90$^{\rm o}$, respectively, which turns out to show a strong correlation with the rapid increases of $\Delta R_{\rm L}^{2\omega}$ (solid black squares) and $\Delta R_{\rm T}^{2\omega}$ (solid red circles). This is in big contrast to the minor decrease of resistivity ($\rho_{\rm L}$) from about 13.1 to 10.3 $\mu\Omega$cm as $T$ goes from 10 to 2 K. Therefore, the rapid increases of the second harmonic signals of $\Delta R_{\rm T}^{2\omega}$ and $\Delta R_{\rm L}^{2\omega}$ below 10 K (Fig. \ref{NLHIdep}(c)) are unlikely scaled with the bulk Drude electron lifetime. Instead, it signifies a crossover to a surface dominant charge transport with inversion symmetry breaking below 10 K, and thus the observed nonlinear Hall signals is likely originated from the surface states in a SRO thin film. In addition, the observation of a large NRTE of $\Delta R_{\rm L}^{2\omega}$ in the longitudinal channel is intriguing, and its amplitude also grows with decreasing $T$ below 10 K, suggesting an intimate relation with the appearance of the nonlinear Hall signals of $\Delta R_{\rm T}^{2\omega}$. However, as demonstrated in Fig. \ref{NLH}(c), the $\alpha$ dependence reveals a clear orthogonality in the $\Delta R_{\rm L}^{2\omega}$ and  $\Delta R_{\rm T}^{2\omega}$. 

We thus propose a real space scenario as illustrated in Fig. \ref{cal}(b), where a $\vec{D}$ along [1\=10]$_{\rm o}$ is accompanied by 1D chiral edge modes along the orthogonal direction of [001]$_{\rm o}$ (orange line). Figure \ref{cal}(c) illustrates a minimum Weyl model with one pair of Weyl nodes with chiral charges of +1 and -1. For the yellow-shaded slice between Weyl node pair of opposite chiral charges, the integration of the total Berry flux across each 2D slice will give a Chern number of 1 accompanied by a unique 1D chiral edge modes at the boundary of the system as shown in the upper panel of Fig. \ref{cal}(c) \cite{Wan2011,Armitage2018}. On the other hand, for green-shaded slice with the Weyl-node pair on the same side, the total Chern number is then zero without the presence of chiral edge modes. The Fermi-arc surface states are thus the zero energy chiral edge modes, connecting the non-overlapped Weyl-node pair on a surface Brillouin zone. By searching for Weyl nodes among four bands of 87 to 90 near the Fermi surface within an energy window of $|\varepsilon-\varepsilon_{\rm F}| \leq$ 30 meV in the calculated SRO band structure, a number of Weyl nodes from band pair (89/90) can be identified and projected on (110)$_{\rm o}$ plane as demonstrated in Fig. \ref{cal}(d) (see Supplementary Note 6). Closed and open triangles correspond to Weyl nodes from two different magnetic easy axis conditions of $M\parallel [110]_{\rm o}$ and -20$^{\rm o}$ $M$ tilting from [110]$_{\rm o}$ toward [1\=10]$_{\rm o}$, respectively (see also Supplementary Note 2). The red and blue colors represent the corresponding chiral charge of +1 and -1, respectively. We note that the yellow-shaded region in Fig. \ref{cal}(d) highlights the non-zero total Chern number and thus supports the presence of 1D chiral edge modes along $k_{\rm z}$. 

When flipping the magnetization in SRO, the signs of the chiral charges also reverse due to the swapping of spin subbands, and both the directions of BCD $\vec{D}$ and 1D chiral edge modes will reverse accordingly. Such 1D chiral edge modes are equivalent to the 1D chiral edge modes in a magnetic TI with quantum anomalous Hall phase \cite{Haldane1988,Chang2013,Kou2014,Checkelsky2014}, where a large NRTE in the longitudinal channel has been recently reported arising from the asymmetric scattering between the 1D chiral edge modes and other surface states \cite{Yasuda2020}. For the Weyl metal SRO, in principle, similar NRTE in the longitudinal channel for bias current along [001]$_{\rm o}$ ($\Delta R_{\rm L}^{2\omega}$ for $\alpha$ = 0$^{\rm o}$ and 180$^{\rm o}$) can thus appear due to the asymmetric scattering between the 1D chiral edge modes and the Fermi-arc surface states. This may also explain the vanishing of $\Delta R_{\rm L}^{2\omega}$ for $\alpha$ = 90$^{\rm o}$ and thus the intriguing orthogonal relation between $\Delta R_{\rm L}^{2\omega}$ and $\Delta R_{\rm T}^{2\omega}$ shown in Fig. \ref{NLH}(c). We note that our observed $\Delta R_{\rm T}^{2\omega}$ due to an effective BCD of surface states may be related to a recently proposed theory \cite{Waw2021} that a hotline with divergent Berry curvature, separating the Fermi-arc surface states and 3D bulk states, may lead to a large nonlinear Hall response. However, the issues regarding the contribution of Fermi-arc surface states to NRTE and nonlinear Hall effect call for more theoretical and experimental efforts.     

\section{Conclusions}
In summary, large nonlinear and nonreciprocal charge transport effects along the longitudinal ($\Delta R_{\rm L}^{2\omega}$) and transverse ($\Delta R_{\rm T}^{2\omega}$) channels were discovered below 10 K in a sunbeam device fabricated from an untwinned thin film SRO grown on miscut STO (001) substrate. Below 10 K, the crossover of weak field transverse MR behavior and also the rapid rise of 2D-like quantum oscillation amplitude not only support the surface dominant charge transport but also agree well with the observed $T$ dependent $\Delta R_{\rm L(T)}^{2\omega}$. The detailed bias current direction dependence reveals an intriguing orthogonality between the observed $\Delta R_{\rm L}^{2\omega}$ and $\Delta R_{\rm T}^{2\omega}$, and, for bias current along [1\=10]$_{\rm o}$ ($\alpha$ = 90$^{\rm o}$), $\Delta R_{\rm T}^{2\omega}$ is at maximum while $\Delta R_{\rm L}^{2\omega}$ is vanishingly small. Considering the dominant roles of the intrinsic AHE and surface charge transport at low temperatures in thin films of SRO/STO system, a scenario of an effective BCD $\vec{D}$ from surface states along [1\=10]$_{\rm o}$ accompanied by 1D chiral edge modes along [001]$_{\rm o}$ was proposed to give a qualitative explanation for the observed $\alpha$ dependent $\Delta R_{\rm L}^{2\omega}$ and $\Delta R_{\rm T}^{2\omega}$, which is supported by the calculated band dispersion with tilted Weyl nodes. Our findings demonstrate the feasibility of using the nonlinear and nonreciprocal charge transport effect as a probe for intriguing topology-related electronic properties in a topological system, such as the BCD from nonlinear Hall and 1D chiral edge modes from NRTE. On the other hand, our observations of NRTE and nonlinear Hall effect in SRO/STO may also highlight the intriguing possibility of investigating surface dominant charge transport behavior in topological thin film systems.          

\textit{Note added in proof.} We recently became aware of a recent theoretical work on NRTE and nonlinear Hall effect due to quantum metric dipole in a special PT-symmetric topological antiferromagnet MnBi2Te4 \cite{Yan2024}. This system is fundamentally different from the PT-asymmetric ferromagnetic Weyl SRO system discussed in current manuscript.

\section*{Methods}
The sunbeam device was patterned on a SRO/STO thin film with SRO layer thickness $t \approx$ 13.7 nm, using standard photolithography followed by argon ion milling. It comprises of 16 Hall bars with $\alpha$ ranging from 0$^{\rm o}$ to 360$^{\rm o}$, and the angle difference between adjacent Hall bars is 22.5$^{\rm o}$. One of the Hall bars was carefully aligned along the SRO orthorhombic [001]$_{\rm o}$ direction, which is defined as $\alpha$ = 0$^{\rm o}$. Each Hall bar has exactly the same geometry with a width of 150 $\mu$m and a length of 290 $\mu$m between longitudinal voltage leads. The Au (35 nm)/Ti (10 nm) electrodes were deposited and fabricated via a subsequent step of photolithography.

The magnetization measurements on SRO/STO thin films were carried out using a SQUID-MPMS system from Quantum Design. The longitudinal (transverse) $\Delta R_{\rm L (T)}^{\omega}$ and $\Delta R_{\rm L(T)}^{2\omega}$ signals were measured simultaneously by a lock-in amplifier at first and second harmonic references, respectively. Rotational anisotropy (RA) SHG measurements were performed using a high-speed rotating scattering plane method described elsewhere \cite{Harter2015}. The light source was a Ti:sapphire laser of central wavelength of 800 nm. The incident beam was focused onto the sample surface at oblique incidence ($\theta$ =  10$^{\rm o}$) with a spot size of $\sim$ 30 $\mu$m. 

The electronic structure calculations for the orthorhombic and monoclinic structures of SrRuO$_3$ were performed using 
the projector augmented-wave method \cite{Kresse}, as implemented in the Vienna $ab$-initio Simulation package \cite{vasp}, 
within the generalized gradient approximation schemes \cite{PBE}. A 9 $\times$ 9 $\times$ 6 Gamma centered $k$-point mesh was used, together with a cutoff energy of 500 eV. The total energy convergence criterion was 10$^{-6}$ eV. The spin-orbit coupling was included in the self-consistent calculations. In the orthorhombic case, the magnetization direction (M) is chosen along $[110]_{\rm o}$. In the monoclinic case, two magnetization directions were considered, namely, M is along $[110]_{\rm o}$ and M is tilted -20$^{o}$ from $[110]_{\rm o}$ towards [1\=10]$_{\rm o}$. In all the cases, the effect of the on-site electronic correlations in the Ru $d$ states (4d$^4$ for Ru$^{4+}$ ) was taken into account by using the rotationally invariant GGA+U scheme \cite{Liechtenstein} with $U$ = 3.0 eV and $J$ = 0.6 eV. We have used Ru $d$-orbitals and O $p$-orbitals to construct the Wannier functions \cite{Marzari,Mostofi} with VASP2WANNIER90 \cite{Franchini} interface. We have used WannierTools \cite{QuanSheng} to search the Weyl points and to identify the chirality of each Weyl point. We have used FINDSYM \cite{Stokes} to identify the magnetic space group symmetry in all the cases with the magnetic moments of both O and Ru atoms included.

   
\section*{Data Availability}
All the supporting data are included in the main text and supplementary information. The data that support the findings of this study are available from corresponding authors upon reasonable request.

\section*{Code Availability}
The input files for the {\it ab initio} calculations using VASP, Wannier tight binding and WannierTools are available upon reasonable request. 

\section*{Acknowledgements}
We thank N. Nagaosa, Q. Niu, Q. Ma, J. Shan, and M. Orlita for valuable discussions. This work was supported by the National Science and Technology Council of Taiwan (NSTC Grants No. 108-2628-M-001-007-MY3,
No. 110-2112-M-002-030-MY3, and No. 111-2112-M001-056-MY3) and the joint project of Academia Sinica and National Taiwan University (Grant No. AS-NTU-110-10).

\section*{Competing interests} The authors declare no competing financial or non-financial interests.

\section*{Author Contributions}
U.K., E.C.H.L., C.T.C., IC.C., and W.L.L. carried out the low-temperature magneto-transport measurements and data analyses. U.K. and A.K.S. grew the epitaxial SRO films. A.K.S., S.Y., C.Y.L., and C.H.H. performed the X-ray measurements at NSRRC in Taiwan. P.V.S.R., G.Y.G., and W.C.L. performed the {\it ab initio} calculations.. Y.J.H., X.W.L., and D.H. performed the SHG measurements and analyses. G.Y.G. and W.L.L. designed the experiment and wrote the manuscript.  



\begin{thebibliography}{69}%
\makeatletter
\providecommand \@ifxundefined [1]{%
 \@ifx{#1\undefined}
}%
\providecommand \@ifnum [1]{%
 \ifnum #1\expandafter \@firstoftwo
 \else \expandafter \@secondoftwo
 \fi
}%
\providecommand \@ifx [1]{%
 \ifx #1\expandafter \@firstoftwo
 \else \expandafter \@secondoftwo
 \fi
}%
\providecommand \natexlab [1]{#1}%
\providecommand \enquote  [1]{``#1''}%
\providecommand \bibnamefont  [1]{#1}%
\providecommand \bibfnamefont [1]{#1}%
\providecommand \citenamefont [1]{#1}%
\providecommand \href@noop [0]{\@secondoftwo}%
\providecommand \href [0]{\begingroup \@sanitize@url \@href}%
\providecommand \@href[1]{\@@startlink{#1}\@@href}%
\providecommand \@@href[1]{\endgroup#1\@@endlink}%
\providecommand \@sanitize@url [0]{\catcode `\\12\catcode `\$12\catcode
  `\&12\catcode `\#12\catcode `\^12\catcode `\_12\catcode `\%12\relax}%
\providecommand \@@startlink[1]{}%
\providecommand \@@endlink[0]{}%
\providecommand \url  [0]{\begingroup\@sanitize@url \@url }%
\providecommand \@url [1]{\endgroup\@href {#1}{\urlprefix }}%
\providecommand \urlprefix  [0]{URL }%
\providecommand \Eprint [0]{\href }%
\providecommand \doibase [0]{https://doi.org/}%
\providecommand \selectlanguage [0]{\@gobble}%
\providecommand \bibinfo  [0]{\@secondoftwo}%
\providecommand \bibfield  [0]{\@secondoftwo}%
\providecommand \translation [1]{[#1]}%
\providecommand \BibitemOpen [0]{}%
\providecommand \bibitemStop [0]{}%
\providecommand \bibitemNoStop [0]{.\EOS\space}%
\providecommand \EOS [0]{\spacefactor3000\relax}%
\providecommand \BibitemShut  [1]{\csname bibitem#1\endcsname}%
\let\auto@bib@innerbib\@empty
\bibitem [{\citenamefont {König}\ \emph {et~al.}(2007)\citenamefont {König},
  \citenamefont {Wiedmann}, \citenamefont {Brüne}, \citenamefont {Roth},
  \citenamefont {Buhmann}, \citenamefont {Molenkamp}, \citenamefont {Qi},\ and\
  \citenamefont {Zhang}}]{Konig2007}%
  \BibitemOpen
  \bibfield  {author} {\bibinfo {author} {\bibfnamefont {M.}~\bibnamefont
  {König}}, \bibinfo {author} {\bibfnamefont {S.}~\bibnamefont {Wiedmann}},
  \bibinfo {author} {\bibfnamefont {C.}~\bibnamefont {Brüne}}, \bibinfo
  {author} {\bibfnamefont {A.}~\bibnamefont {Roth}}, \bibinfo {author}
  {\bibfnamefont {H.}~\bibnamefont {Buhmann}}, \bibinfo {author} {\bibfnamefont
  {L.~W.}\ \bibnamefont {Molenkamp}}, \bibinfo {author} {\bibfnamefont {X.-L.}\
  \bibnamefont {Qi}},\ and\ \bibinfo {author} {\bibfnamefont {S.-C.}\
  \bibnamefont {Zhang}},\ }\bibfield  {title} {\bibinfo {title}\textit{Quantum spin
  {Hall} insulator state in {HgTe} quantum wells},\ }\href@noop {} {\bibfield
  {journal} {\bibinfo  {journal} {Science}\ }\textbf {\bibinfo {volume}
  {318}},\ \bibinfo {pages} {766} (\bibinfo {year} {2007})}\BibitemShut
  {NoStop}%
\bibitem [{\citenamefont {Du}\ \emph {et~al.}(2015)\citenamefont {Du},
  \citenamefont {Knez}, \citenamefont {Sullivan},\ and\ \citenamefont
  {Du}}]{Du2015}%
  \BibitemOpen
  \bibfield  {author} {\bibinfo {author} {\bibfnamefont {L.}~\bibnamefont
  {Du}}, \bibinfo {author} {\bibfnamefont {I.}~\bibnamefont {Knez}}, \bibinfo
  {author} {\bibfnamefont {G.}~\bibnamefont {Sullivan}},\ and\ \bibinfo
  {author} {\bibfnamefont {R.-R.}\ \bibnamefont {Du}},\ }\bibfield  {title}
  {\bibinfo {title} \textit{Robust helical edge transport in gated
  {InAs}/{GaSb} bilayers},\ }\href@noop {} {\bibfield
  {journal} {\bibinfo  {journal} {Phys. Rev. Lett.}\ }\textbf {\bibinfo
  {volume} {114}},\ \bibinfo {pages} {096802} (\bibinfo {year}
  {2015})}\BibitemShut {NoStop}%
\bibitem [{\citenamefont {Fei}\ \emph {et~al.}(2017)\citenamefont {Fei},
  \citenamefont {Palomaki}, \citenamefont {Wu}, \citenamefont {Zhao},
  \citenamefont {Cai}, \citenamefont {Sun}, \citenamefont {Nguyen},
  \citenamefont {Finney}, \citenamefont {Xu},\ and\ \citenamefont
  {Cobden}}]{Fei2017}%
  \BibitemOpen
  \bibfield  {author} {\bibinfo {author} {\bibfnamefont {Z.}~\bibnamefont
  {Fei}}, \bibinfo {author} {\bibfnamefont {T.}~\bibnamefont {Palomaki}},
  \bibinfo {author} {\bibfnamefont {S.}~\bibnamefont {Wu}}, \bibinfo {author}
  {\bibfnamefont {W.}~\bibnamefont {Zhao}}, \bibinfo {author} {\bibfnamefont
  {X.}~\bibnamefont {Cai}}, \bibinfo {author} {\bibfnamefont {B.}~\bibnamefont
  {Sun}}, \bibinfo {author} {\bibfnamefont {P.}~\bibnamefont {Nguyen}},
  \bibinfo {author} {\bibfnamefont {J.}~\bibnamefont {Finney}}, \bibinfo
  {author} {\bibfnamefont {X.}~\bibnamefont {Xu}},\ and\ \bibinfo {author}
  {\bibfnamefont {D.~H.}\ \bibnamefont {Cobden}},\ }\bibfield  {title}
  {\bibinfo {title} \textit{Edge conduction in monolayer {WTe$_2$}},\ }\href@noop {}
  {\bibfield  {journal} {\bibinfo  {journal} {Nat. Physics}\ }\textbf {\bibinfo
  {volume} {13}},\ \bibinfo {pages} {677} (\bibinfo {year} {2017})}\BibitemShut
  {NoStop}%
\bibitem [{\citenamefont {Tang}\ \emph {et~al.}(2017)\citenamefont {Tang},
  \citenamefont {Zhang}, \citenamefont {Wong}, \citenamefont {Pedramrazi},
  \citenamefont {Tsai}, \citenamefont {Jia}, \citenamefont {Moritz},
  \citenamefont {Claassen}, \citenamefont {Ryu}, \citenamefont {Kahn},
  \citenamefont {Jiang}, \citenamefont {Yan}, \citenamefont {Hashimoto},
  \citenamefont {Lu}, \citenamefont {Moore}, \citenamefont {Hwang},
  \citenamefont {Hwang}, \citenamefont {Hussain}, \citenamefont {Chen},
  \citenamefont {Ugeda}, \citenamefont {Liu}, \citenamefont {Xie},
  \citenamefont {Devereaux}, \citenamefont {Crommie}, \citenamefont {Mo},\ and\
  \citenamefont {Shen}}]{Tang2017}%
  \BibitemOpen
  \bibfield  {author} {\bibinfo {author} {\bibfnamefont {S.}~\bibnamefont
  {Tang}}, \bibinfo {author} {\bibfnamefont {C.}~\bibnamefont {Zhang}},
  \bibinfo {author} {\bibfnamefont {D.}~\bibnamefont {Wong}}, \bibinfo {author}
  {\bibfnamefont {Z.}~\bibnamefont {Pedramrazi}}, \bibinfo {author}
  {\bibfnamefont {H.-Z.}\ \bibnamefont {Tsai}}, \bibinfo {author}
  {\bibfnamefont {C.}~\bibnamefont {Jia}}, \bibinfo {author} {\bibfnamefont
  {B.}~\bibnamefont {Moritz}}, \bibinfo {author} {\bibfnamefont
  {M.}~\bibnamefont {Claassen}}, \bibinfo {author} {\bibfnamefont
  {H.}~\bibnamefont {Ryu}}, \bibinfo {author} {\bibfnamefont {S.}~\bibnamefont
  {Kahn}}, \bibinfo {author} {\bibfnamefont {J.}~\bibnamefont {Jiang}},
  \bibinfo {author} {\bibfnamefont {H.}~\bibnamefont {Yan}}, \bibinfo {author}
  {\bibfnamefont {M.}~\bibnamefont {Hashimoto}}, \bibinfo {author}
  {\bibfnamefont {D.}~\bibnamefont {Lu}}, \bibinfo {author} {\bibfnamefont
  {R.~G.}\ \bibnamefont {Moore}}, \bibinfo {author} {\bibfnamefont {C.-C.}\
  \bibnamefont {Hwang}}, \bibinfo {author} {\bibfnamefont {C.}~\bibnamefont
  {Hwang}}, \bibinfo {author} {\bibfnamefont {Z.}~\bibnamefont {Hussain}},
  \bibinfo {author} {\bibfnamefont {Y.}~\bibnamefont {Chen}}, \bibinfo {author}
  {\bibfnamefont {M.~M.}\ \bibnamefont {Ugeda}}, \bibinfo {author}
  {\bibfnamefont {Z.}~\bibnamefont {Liu}}, \bibinfo {author} {\bibfnamefont
  {X.}~\bibnamefont {Xie}}, \bibinfo {author} {\bibfnamefont {T.~P.}\
  \bibnamefont {Devereaux}}, \bibinfo {author} {\bibfnamefont {M.~F.}\
  \bibnamefont {Crommie}}, \bibinfo {author} {\bibfnamefont {S.-K.}\
  \bibnamefont {Mo}},\ and\ \bibinfo {author} {\bibfnamefont {Z.-X.}\
  \bibnamefont {Shen}},\ }\bibfield  {title} {\bibinfo {title} \textit{Quantum spin
  {Hall} state in monolayer {1T'-WTe$_2$}},\ }\href@noop {} {\bibfield
  {journal} {\bibinfo  {journal} {Nat. Physics}\ }\textbf {\bibinfo {volume}
  {13}},\ \bibinfo {pages} {683} (\bibinfo {year} {2017})}\BibitemShut
  {NoStop}%
\bibitem [{\citenamefont {Hsieh}\ \emph {et~al.}(2008)\citenamefont {Hsieh},
  \citenamefont {Qian}, \citenamefont {Wray}, \citenamefont {Xia},
  \citenamefont {Hor}, \citenamefont {Cava},\ and\ \citenamefont
  {Hasan}}]{Hsieh2008}%
  \BibitemOpen
  \bibfield  {author} {\bibinfo {author} {\bibfnamefont {D.}~\bibnamefont
  {Hsieh}}, \bibinfo {author} {\bibfnamefont {D.}~\bibnamefont {Qian}},
  \bibinfo {author} {\bibfnamefont {L.}~\bibnamefont {Wray}}, \bibinfo {author}
  {\bibfnamefont {Y.}~\bibnamefont {Xia}}, \bibinfo {author} {\bibfnamefont
  {Y.~S.}\ \bibnamefont {Hor}}, \bibinfo {author} {\bibfnamefont {R.~J.}\
  \bibnamefont {Cava}},\ and\ \bibinfo {author} {\bibfnamefont {M.~Z.}\
  \bibnamefont {Hasan}},\ }\bibfield  {title} {\bibinfo {title} \textit{A topological
  {Dirac} insulator in a quantum spin {Hall} phase},\ }\href@noop {} {\bibfield
   {journal} {\bibinfo  {journal} {Nature}\ }\textbf {\bibinfo {volume} {452}},\
  \bibinfo {pages} {970} (\bibinfo {year} {2008})}\BibitemShut {NoStop}%
\bibitem [{\citenamefont {Alpichshev}\ \emph {et~al.}(2010)\citenamefont
  {Alpichshev}, \citenamefont {Analytis}, \citenamefont {Chu}, \citenamefont
  {Fisher}, \citenamefont {Chen}, \citenamefont {Shen}, \citenamefont {Fang},\
  and\ \citenamefont {Kapitulnik}}]{Alpi2010}%
  \BibitemOpen
  \bibfield  {author} {\bibinfo {author} {\bibfnamefont {Z.}~\bibnamefont
  {Alpichshev}}, \bibinfo {author} {\bibfnamefont {J.~G.}\ \bibnamefont
  {Analytis}}, \bibinfo {author} {\bibfnamefont {J.-H.}\ \bibnamefont {Chu}},
  \bibinfo {author} {\bibfnamefont {I.~R.}\ \bibnamefont {Fisher}}, \bibinfo
  {author} {\bibfnamefont {Y.~L.}\ \bibnamefont {Chen}}, \bibinfo {author}
  {\bibfnamefont {Z.~X.}\ \bibnamefont {Shen}}, \bibinfo {author}
  {\bibfnamefont {A.}~\bibnamefont {Fang}},\ and\ \bibinfo {author}
  {\bibfnamefont {A.}~\bibnamefont {Kapitulnik}},\ }\bibfield  {title}
  {\bibinfo {title} \textit{{STM} imaging of electronic waves on the surface of
  {Bi$_{2}$Te$_{3}$}: Topologically protected surface states and hexagonal
  warping effects},\ }\href@noop {} {\bibfield  {journal} {\bibinfo  {journal}
  {Phys. Rev. Lett.}\ }\textbf {\bibinfo {volume} {104}},\ \bibinfo {pages}
  {016401} (\bibinfo {year} {2010})}\BibitemShut {NoStop}%
\bibitem [{\citenamefont {Hasan}\ and\ \citenamefont {Kane}(2010)}]{Hasan2010}%
  \BibitemOpen
  \bibfield  {author} {\bibinfo {author} {\bibfnamefont {M.~Z.}\ \bibnamefont
  {Hasan}}\ and\ \bibinfo {author} {\bibfnamefont {C.~L.}\ \bibnamefont
  {Kane}},\ }\bibfield  {title} {\bibinfo {title} \textit{Colloquium: Topological
  insulators},\ }\href@noop {} {\bibfield  {journal} {\bibinfo  {journal} {Rev.
  Mod. Phys.}\ }\textbf {\bibinfo {volume} {82}},\ \bibinfo {pages} {3045}
  (\bibinfo {year} {2010})}\BibitemShut {NoStop}%
\bibitem [{\citenamefont {Chang}\ \emph {et~al.}(2013)\citenamefont {Chang},
  \citenamefont {Zhang}, \citenamefont {Feng}, \citenamefont {Shen},
  \citenamefont {Zhang}, \citenamefont {Guo}, \citenamefont {Li}, \citenamefont
  {Ou}, \citenamefont {Wei}, \citenamefont {Wang}, \citenamefont {Ji},
  \citenamefont {Feng}, \citenamefont {Ji}, \citenamefont {Chen}, \citenamefont
  {Jia}, \citenamefont {Dai}, \citenamefont {Fang}, \citenamefont {Zhang},
  \citenamefont {He}, \citenamefont {Wang}, \citenamefont {Lu}, \citenamefont
  {Ma},\ and\ \citenamefont {Xue}}]{Chang2013}%
  \BibitemOpen
  \bibfield  {author} {\bibinfo {author} {\bibfnamefont {C.-Z.}\ \bibnamefont
  {Chang}}, \bibinfo {author} {\bibfnamefont {J.}~\bibnamefont {Zhang}},
  \bibinfo {author} {\bibfnamefont {X.}~\bibnamefont {Feng}}, \bibinfo {author}
  {\bibfnamefont {J.}~\bibnamefont {Shen}}, \bibinfo {author} {\bibfnamefont
  {Z.}~\bibnamefont {Zhang}}, \bibinfo {author} {\bibfnamefont
  {M.}~\bibnamefont {Guo}}, \bibinfo {author} {\bibfnamefont {K.}~\bibnamefont
  {Li}}, \bibinfo {author} {\bibfnamefont {Y.}~\bibnamefont {Ou}}, \bibinfo
  {author} {\bibfnamefont {P.}~\bibnamefont {Wei}}, \bibinfo {author}
  {\bibfnamefont {L.-L.}\ \bibnamefont {Wang}}, \bibinfo {author}
  {\bibfnamefont {Z.-Q.}\ \bibnamefont {Ji}}, \bibinfo {author} {\bibfnamefont
  {Y.}~\bibnamefont {Feng}}, \bibinfo {author} {\bibfnamefont {S.}~\bibnamefont
  {Ji}}, \bibinfo {author} {\bibfnamefont {X.}~\bibnamefont {Chen}}, \bibinfo
  {author} {\bibfnamefont {J.}~\bibnamefont {Jia}}, \bibinfo {author}
  {\bibfnamefont {X.}~\bibnamefont {Dai}}, \bibinfo {author} {\bibfnamefont
  {Z.}~\bibnamefont {Fang}}, \bibinfo {author} {\bibfnamefont {S.-C.}\
  \bibnamefont {Zhang}}, \bibinfo {author} {\bibfnamefont {K.}~\bibnamefont
  {He}}, \bibinfo {author} {\bibfnamefont {Y.}~\bibnamefont {Wang}}, \bibinfo
  {author} {\bibfnamefont {L.}~\bibnamefont {Lu}}, \bibinfo {author}
  {\bibfnamefont {X.-C.}\ \bibnamefont {Ma}},\ and\ \bibinfo {author}
  {\bibfnamefont {Q.-K.}\ \bibnamefont {Xue}},\ }\bibfield  {title} {\bibinfo
  {title} \textit{Experimental observation of the quantum anomalous {Hall} effect in a
  magnetic topological insulator},\ }\href@noop {} {\bibfield  {journal}
  {\bibinfo  {journal} {Science}\ }\textbf {\bibinfo {volume} {340}},\ \bibinfo
  {pages} {167} (\bibinfo {year} {2013})}\BibitemShut {NoStop}%
\bibitem [{\citenamefont {Kou}\ \emph {et~al.}(2014)\citenamefont {Kou},
  \citenamefont {Guo}, \citenamefont {Fan}, \citenamefont {Pan}, \citenamefont
  {Lang}, \citenamefont {Jiang}, \citenamefont {Shao}, \citenamefont {Nie},
  \citenamefont {Murata}, \citenamefont {Tang}, \citenamefont {Wang},
  \citenamefont {He}, \citenamefont {Lee}, \citenamefont {Lee},\ and\
  \citenamefont {Wang}}]{Kou2014}%
  \BibitemOpen
  \bibfield  {author} {\bibinfo {author} {\bibfnamefont {X.}~\bibnamefont
  {Kou}}, \bibinfo {author} {\bibfnamefont {S.-T.}\ \bibnamefont {Guo}},
  \bibinfo {author} {\bibfnamefont {Y.}~\bibnamefont {Fan}}, \bibinfo {author}
  {\bibfnamefont {L.}~\bibnamefont {Pan}}, \bibinfo {author} {\bibfnamefont
  {M.}~\bibnamefont {Lang}}, \bibinfo {author} {\bibfnamefont {Y.}~\bibnamefont
  {Jiang}}, \bibinfo {author} {\bibfnamefont {Q.}~\bibnamefont {Shao}},
  \bibinfo {author} {\bibfnamefont {T.}~\bibnamefont {Nie}}, \bibinfo {author}
  {\bibfnamefont {K.}~\bibnamefont {Murata}}, \bibinfo {author} {\bibfnamefont
  {J.}~\bibnamefont {Tang}}, \bibinfo {author} {\bibfnamefont {Y.}~\bibnamefont
  {Wang}}, \bibinfo {author} {\bibfnamefont {L.}~\bibnamefont {He}}, \bibinfo
  {author} {\bibfnamefont {T.-K.}\ \bibnamefont {Lee}}, \bibinfo {author}
  {\bibfnamefont {W.-L.}\ \bibnamefont {Lee}},\ and\ \bibinfo {author}
  {\bibfnamefont {K.~L.}\ \bibnamefont {Wang}},\ }\bibfield  {title} {\bibinfo
  {title} \textit{Scale-invariant quantum anomalous {Hall} effect in magnetic
  topological insulators beyond the two-dimensional limit},\ }\href@noop {}
  {\bibfield  {journal} {\bibinfo  {journal} {Phys. Rev. Lett.}\ }\textbf
  {\bibinfo {volume} {113}},\ \bibinfo {pages} {137201} (\bibinfo {year}
  {2014})}\BibitemShut {NoStop}%
\bibitem [{\citenamefont {Checkelsky}\ \emph {et~al.}(2014)\citenamefont
  {Checkelsky}, \citenamefont {Yoshimi}, \citenamefont {Tsukazaki},
  \citenamefont {Takahashi}, \citenamefont {Kozuka}, \citenamefont {Falson},
  \citenamefont {Kawasaki},\ and\ \citenamefont {Tokura}}]{Checkelsky2014}%
  \BibitemOpen
  \bibfield  {author} {\bibinfo {author} {\bibfnamefont {J.~G.}\ \bibnamefont
  {Checkelsky}}, \bibinfo {author} {\bibfnamefont {R.}~\bibnamefont {Yoshimi}},
  \bibinfo {author} {\bibfnamefont {A.}~\bibnamefont {Tsukazaki}}, \bibinfo
  {author} {\bibfnamefont {K.~S.}\ \bibnamefont {Takahashi}}, \bibinfo {author}
  {\bibfnamefont {Y.}~\bibnamefont {Kozuka}}, \bibinfo {author} {\bibfnamefont
  {J.}~\bibnamefont {Falson}}, \bibinfo {author} {\bibfnamefont
  {M.}~\bibnamefont {Kawasaki}},\ and\ \bibinfo {author} {\bibfnamefont
  {Y.}~\bibnamefont {Tokura}},\ }\bibfield  {title} {\bibinfo {title}
  \textit{Trajectory of the anomalous {Hall} effect towards the quantized state in a
  ferromagnetic topological insulator},\ }\href@noop {} {\bibfield  {journal}
  {\bibinfo  {journal} {Nat. Physics}\ }\textbf {\bibinfo {volume} {10}},\
  \bibinfo {pages} {731} (\bibinfo {year} {2014})}\BibitemShut {NoStop}%
\bibitem [{\citenamefont {Haldane}(1988)}]{Haldane1988}%
  \BibitemOpen
  \bibfield  {author} {\bibinfo {author} {\bibfnamefont {F.~D.~M.}\
  \bibnamefont {Haldane}},\ }\bibfield  {title} {\bibinfo {title} \textit{Model for a
  quantum {Hall} effect without {Landau} levels: Condensed-matter realization
  of the "parity anomaly"},\ }\href@noop {} {\bibfield  {journal} {\bibinfo
  {journal} {Phys. Rev. Lett.}\ }\textbf {\bibinfo {volume} {61}},\ \bibinfo
  {pages} {2015} (\bibinfo {year} {1988})}\BibitemShut {NoStop}%
\bibitem [{\citenamefont {Wan}\ \emph {et~al.}(2011)\citenamefont {Wan},
  \citenamefont {Turner}, \citenamefont {Vishwanath},\ and\ \citenamefont
  {Savrasov}}]{Wan2011}%
  \BibitemOpen
  \bibfield  {author} {\bibinfo {author} {\bibfnamefont {X.}~\bibnamefont
  {Wan}}, \bibinfo {author} {\bibfnamefont {A.~M.}\ \bibnamefont {Turner}},
  \bibinfo {author} {\bibfnamefont {A.}~\bibnamefont {Vishwanath}},\ and\
  \bibinfo {author} {\bibfnamefont {S.~Y.}\ \bibnamefont {Savrasov}},\
  }\bibfield  {title} {\bibinfo {title} \textit{Topological semimetal and {Fermi-arc}
  surface states in the electronic structure of pyrochlore iridates},\
  }\href@noop {} {\bibfield  {journal} {\bibinfo  {journal} {Phys. Rev. B}\
  }\textbf {\bibinfo {volume} {83}},\ \bibinfo {pages} {205101} (\bibinfo
  {year} {2011})}\BibitemShut {NoStop}%
\bibitem [{\citenamefont {Wang}\ \emph {et~al.}(2012)\citenamefont {Wang},
  \citenamefont {Sun}, \citenamefont {Chen}, \citenamefont {Franchini},
  \citenamefont {Xu}, \citenamefont {Weng}, \citenamefont {Dai},\ and\
  \citenamefont {Fang}}]{Wang2012}%
  \BibitemOpen
  \bibfield  {author} {\bibinfo {author} {\bibfnamefont {Z.}~\bibnamefont
  {Wang}}, \bibinfo {author} {\bibfnamefont {Y.}~\bibnamefont {Sun}}, \bibinfo
  {author} {\bibfnamefont {X.-Q.}\ \bibnamefont {Chen}}, \bibinfo {author}
  {\bibfnamefont {C.}~\bibnamefont {Franchini}}, \bibinfo {author}
  {\bibfnamefont {G.}~\bibnamefont {Xu}}, \bibinfo {author} {\bibfnamefont
  {H.}~\bibnamefont {Weng}}, \bibinfo {author} {\bibfnamefont {X.}~\bibnamefont
  {Dai}},\ and\ \bibinfo {author} {\bibfnamefont {Z.}~\bibnamefont {Fang}},\
  }\bibfield  {title} {\bibinfo {title} \textit{Dirac semimetal and topological phase
  transitions in {A$_{3}$Bi} ({A = Na, K, Rb})},\ }\href@noop {} {\bibfield
  {journal} {\bibinfo  {journal} {Phys. Rev. B}\ }\textbf {\bibinfo {volume}
  {85}},\ \bibinfo {pages} {195320} (\bibinfo {year} {2012})}\BibitemShut
  {NoStop}%
\bibitem [{\citenamefont {Liang}\ \emph {et~al.}(2015)\citenamefont {Liang},
  \citenamefont {Gibson}, \citenamefont {Ali}, \citenamefont {Liu},
  \citenamefont {Cava},\ and\ \citenamefont {Ong}}]{Liang2015}%
  \BibitemOpen
  \bibfield  {author} {\bibinfo {author} {\bibfnamefont {T.}~\bibnamefont
  {Liang}}, \bibinfo {author} {\bibfnamefont {Q.}~\bibnamefont {Gibson}},
  \bibinfo {author} {\bibfnamefont {M.~N.}\ \bibnamefont {Ali}}, \bibinfo
  {author} {\bibfnamefont {M.}~\bibnamefont {Liu}}, \bibinfo {author}
  {\bibfnamefont {R.~J.}\ \bibnamefont {Cava}},\ and\ \bibinfo {author}
  {\bibfnamefont {N.~P.}\ \bibnamefont {Ong}},\ }\bibfield  {title} {\bibinfo
  {title} \textit{Ultrahigh mobility and giant magnetoresistance in the {Dirac}
  semimetal {Cd$_{3}$As$_{2}$}},\ }\href@noop {} {\bibfield  {journal}
  {\bibinfo  {journal} {Nat. Mater.}\ }\textbf {\bibinfo {volume} {14}},\
  \bibinfo {pages} {280} (\bibinfo {year} {2015})}\BibitemShut {NoStop}%
\bibitem [{\citenamefont {Huang}\ \emph {et~al.}(2015)\citenamefont {Huang},
  \citenamefont {Zhao}, \citenamefont {Long}, \citenamefont {Wang},
  \citenamefont {Chen}, \citenamefont {Yang}, \citenamefont {Liang},
  \citenamefont {Xue}, \citenamefont {Weng}, \citenamefont {Fang},
  \citenamefont {Dai},\ and\ \citenamefont {Chen}}]{Huang2015}%
  \BibitemOpen
  \bibfield  {author} {\bibinfo {author} {\bibfnamefont {X.}~\bibnamefont
  {Huang}}, \bibinfo {author} {\bibfnamefont {L.}~\bibnamefont {Zhao}},
  \bibinfo {author} {\bibfnamefont {Y.}~\bibnamefont {Long}}, \bibinfo {author}
  {\bibfnamefont {P.}~\bibnamefont {Wang}}, \bibinfo {author} {\bibfnamefont
  {D.}~\bibnamefont {Chen}}, \bibinfo {author} {\bibfnamefont {Z.}~\bibnamefont
  {Yang}}, \bibinfo {author} {\bibfnamefont {H.}~\bibnamefont {Liang}},
  \bibinfo {author} {\bibfnamefont {M.}~\bibnamefont {Xue}}, \bibinfo {author}
  {\bibfnamefont {H.}~\bibnamefont {Weng}}, \bibinfo {author} {\bibfnamefont
  {Z.}~\bibnamefont {Fang}}, \bibinfo {author} {\bibfnamefont {X.}~\bibnamefont
  {Dai}},\ and\ \bibinfo {author} {\bibfnamefont {G.}~\bibnamefont {Chen}},\
  }\bibfield  {title} {\bibinfo {title} \textit{Observation of the
  chiral-anomaly-induced negative magnetoresistance in {3D} {Weyl} semimetal
  {TaAs}},\ }\href@noop {} {\bibfield  {journal} {\bibinfo  {journal} {Phys.
  Rev. X}\ }\textbf {\bibinfo {volume} {5}},\ \bibinfo {pages} {031023}
  (\bibinfo {year} {2015})}\BibitemShut {NoStop}%
\bibitem [{\citenamefont {Xiong}\ \emph {et~al.}(2015)\citenamefont {Xiong},
  \citenamefont {Kushwaha}, \citenamefont {Liang}, \citenamefont {Krizan},
  \citenamefont {Hirschberger}, \citenamefont {Wang}, \citenamefont {Cava},\
  and\ \citenamefont {Ong}}]{Xiong2015}%
  \BibitemOpen
  \bibfield  {author} {\bibinfo {author} {\bibfnamefont {J.}~\bibnamefont
  {Xiong}}, \bibinfo {author} {\bibfnamefont {S.~K.}\ \bibnamefont {Kushwaha}},
  \bibinfo {author} {\bibfnamefont {T.}~\bibnamefont {Liang}}, \bibinfo
  {author} {\bibfnamefont {J.~W.}\ \bibnamefont {Krizan}}, \bibinfo {author}
  {\bibfnamefont {M.}~\bibnamefont {Hirschberger}}, \bibinfo {author}
  {\bibfnamefont {W.}~\bibnamefont {Wang}}, \bibinfo {author} {\bibfnamefont
  {R.~J.}\ \bibnamefont {Cava}},\ and\ \bibinfo {author} {\bibfnamefont
  {N.~P.}\ \bibnamefont {Ong}},\ }\bibfield  {title} {\bibinfo {title}
  \textit{Evidence for the chiral anomaly in the {Dirac} semimetal {Na$_3$Bi}},\
  }\href@noop {} {\bibfield  {journal} {\bibinfo  {journal} {Science}\ }\textbf
  {\bibinfo {volume} {350}},\ \bibinfo {pages} {413} (\bibinfo {year}
  {2015})}\BibitemShut {NoStop}%
\bibitem [{\citenamefont {Armitage}\ \emph {et~al.}(2018)\citenamefont
  {Armitage}, \citenamefont {Mele},\ and\ \citenamefont
  {Vishwanath}}]{Armitage2018}%
  \BibitemOpen
  \bibfield  {author} {\bibinfo {author} {\bibfnamefont {N.~P.}\ \bibnamefont
  {Armitage}}, \bibinfo {author} {\bibfnamefont {E.~J.}\ \bibnamefont {Mele}},\
  and\ \bibinfo {author} {\bibfnamefont {A.}~\bibnamefont {Vishwanath}},\
  }\bibfield  {title} {\bibinfo {title} \textit{{Weyl} and {Dirac} semimetals in
  three-dimensional solids},\ }\href@noop {} {\bibfield  {journal} {\bibinfo
  {journal} {Rev. Mod. Phys.}\ }\textbf {\bibinfo {volume} {90}},\ \bibinfo
  {pages} {015001} (\bibinfo {year} {2018})}\BibitemShut {NoStop}%
\bibitem [{\citenamefont {Potter}\ \emph {et~al.}(2014)\citenamefont {Potter},
  \citenamefont {Kimchi},\ and\ \citenamefont {Vishwanath}}]{Potter2014}%
  \BibitemOpen
  \bibfield  {author} {\bibinfo {author} {\bibfnamefont {A.~C.}\ \bibnamefont
  {Potter}}, \bibinfo {author} {\bibfnamefont {I.}~\bibnamefont {Kimchi}},\
  and\ \bibinfo {author} {\bibfnamefont {A.}~\bibnamefont {Vishwanath}},\
  }\bibfield  {title} {\bibinfo {title} \textit{Quantum oscillations from surface
  {Fermi} arcs in {Weyl} and {Dirac} semimetals},\ }\href@noop {} {\bibfield
  {journal} {\bibinfo  {journal} {Nat. Commun.}\ }\textbf {\bibinfo {volume}
  {5}},\ \bibinfo {pages} {5161} (\bibinfo {year} {2014})}\BibitemShut
  {NoStop}%
\bibitem [{\citenamefont {Wawrzik}\ \emph {et~al.}(2021)\citenamefont
  {Wawrzik}, \citenamefont {You}, \citenamefont {Facio}, \citenamefont {van~den
  Brink},\ and\ \citenamefont {Sodemann}}]{Waw2021}%
  \BibitemOpen
  \bibfield  {author} {\bibinfo {author} {\bibfnamefont {D.}~\bibnamefont
  {Wawrzik}}, \bibinfo {author} {\bibfnamefont {J.-S.}\ \bibnamefont {You}},
  \bibinfo {author} {\bibfnamefont {J.~I.}\ \bibnamefont {Facio}}, \bibinfo
  {author} {\bibfnamefont {J.}~\bibnamefont {van~den Brink}},\ and\ \bibinfo
  {author} {\bibfnamefont {I.}~\bibnamefont {Sodemann}},\ }\bibfield  {title}
  {\bibinfo {title} \textit{Infinite {Berry} curvature of {Weyl} {Fermi} arcs},\
  }\href@noop {} {\bibfield  {journal} {\bibinfo  {journal} {Phys. Rev. Lett.}\
  }\textbf {\bibinfo {volume} {127}},\ \bibinfo {pages} {056601} (\bibinfo
  {year} {2021})}\BibitemShut {NoStop}%
\bibitem [{\citenamefont {Gao}\ \emph {et~al.}(2014)\citenamefont {Gao},
  \citenamefont {Yang},\ and\ \citenamefont {Niu}}]{Gao2014}%
  \BibitemOpen
  \bibfield  {author} {\bibinfo {author} {\bibfnamefont {Y.}~\bibnamefont
  {Gao}}, \bibinfo {author} {\bibfnamefont {S.~A.}\ \bibnamefont {Yang}},\ and\
  \bibinfo {author} {\bibfnamefont {Q.}~\bibnamefont {Niu}},\ }\bibfield
  {title} {\bibinfo {title} \textit{Field induced positional shift of {Bloch}
  electrons and its dynamical implications},\ }\href@noop {} {\bibfield
  {journal} {\bibinfo  {journal} {Phys. Rev. Lett.}\ }\textbf {\bibinfo
  {volume} {112}},\ \bibinfo {pages} {166601} (\bibinfo {year}
  {2014})}\BibitemShut {NoStop}%
\bibitem [{\citenamefont {Sodemann}\ and\ \citenamefont
  {Fu}(2015)}]{Sodemann2015}%
  \BibitemOpen
  \bibfield  {author} {\bibinfo {author} {\bibfnamefont {I.}~\bibnamefont
  {Sodemann}}\ and\ \bibinfo {author} {\bibfnamefont {L.}~\bibnamefont {Fu}},\
  }\bibfield  {title} {\bibinfo {title} \textit{Quantum nonlinear {Hall} effect
  induced by {Berry} curvature dipole in time-reversal invariant materials},\
  }\href@noop {} {\bibfield  {journal} {\bibinfo  {journal} {Phys. Rev. Lett.}\
  }\textbf {\bibinfo {volume} {115}},\ \bibinfo {pages} {216806} (\bibinfo
  {year} {2015})}\BibitemShut {NoStop}%
\bibitem [{\citenamefont {Ma}\ \emph {et~al.}(2019)\citenamefont {Ma},
  \citenamefont {Xu}, \citenamefont {Shen}, \citenamefont {MacNeill},
  \citenamefont {Fatemi}, \citenamefont {Chang}, \citenamefont {Mier~Valdivia},
  \citenamefont {Wu}, \citenamefont {Du}, \citenamefont {Hsu}, \citenamefont
  {Fang}, \citenamefont {Gibson}, \citenamefont {Watanabe}, \citenamefont
  {Taniguchi}, \citenamefont {Cava}, \citenamefont {Kaxiras}, \citenamefont
  {Lu}, \citenamefont {Lin}, \citenamefont {Fu}, \citenamefont {Gedik},\ and\
  \citenamefont {Jarillo-Herrero}}]{Ma2019}%
  \BibitemOpen
  \bibfield  {author} {\bibinfo {author} {\bibfnamefont {Q.}~\bibnamefont
  {Ma}}, \bibinfo {author} {\bibfnamefont {S.-Y.}\ \bibnamefont {Xu}}, \bibinfo
  {author} {\bibfnamefont {H.}~\bibnamefont {Shen}}, \bibinfo {author}
  {\bibfnamefont {D.}~\bibnamefont {MacNeill}}, \bibinfo {author}
  {\bibfnamefont {V.}~\bibnamefont {Fatemi}}, \bibinfo {author} {\bibfnamefont
  {T.-R.}\ \bibnamefont {Chang}}, \bibinfo {author} {\bibfnamefont {A.~M.}\
  \bibnamefont {Mier~Valdivia}}, \bibinfo {author} {\bibfnamefont
  {S.}~\bibnamefont {Wu}}, \bibinfo {author} {\bibfnamefont {Z.}~\bibnamefont
  {Du}}, \bibinfo {author} {\bibfnamefont {C.-H.}\ \bibnamefont {Hsu}},
  \bibinfo {author} {\bibfnamefont {S.}~\bibnamefont {Fang}}, \bibinfo {author}
  {\bibfnamefont {Q.~D.}\ \bibnamefont {Gibson}}, \bibinfo {author}
  {\bibfnamefont {K.}~\bibnamefont {Watanabe}}, \bibinfo {author}
  {\bibfnamefont {T.}~\bibnamefont {Taniguchi}}, \bibinfo {author}
  {\bibfnamefont {R.~J.}\ \bibnamefont {Cava}}, \bibinfo {author}
  {\bibfnamefont {E.}~\bibnamefont {Kaxiras}}, \bibinfo {author} {\bibfnamefont
  {H.-Z.}\ \bibnamefont {Lu}}, \bibinfo {author} {\bibfnamefont
  {H.}~\bibnamefont {Lin}}, \bibinfo {author} {\bibfnamefont {L.}~\bibnamefont
  {Fu}}, \bibinfo {author} {\bibfnamefont {N.}~\bibnamefont {Gedik}},\ and\
  \bibinfo {author} {\bibfnamefont {P.}~\bibnamefont {Jarillo-Herrero}},\
  }\bibfield  {title} {\bibinfo {title} \textit{Observation of the nonlinear {Hall}
  effect under time-reversal-symmetric conditions},\ }\href@noop {} {\bibfield
  {journal} {\bibinfo  {journal} {Nature}\ }\textbf {\bibinfo {volume} {565}},\
  \bibinfo {pages} {337} (\bibinfo {year} {2019})}\BibitemShut {NoStop}%
\bibitem [{\citenamefont {Yasuda}\ \emph {et~al.}(2020)\citenamefont {Yasuda},
  \citenamefont {Morimoto}, \citenamefont {Yoshimi}, \citenamefont {Mogi},
  \citenamefont {Tsukazaki}, \citenamefont {Kawamura}, \citenamefont
  {Takahashi}, \citenamefont {Kawasaki}, \citenamefont {Nagaosa},\ and\
  \citenamefont {Tokura}}]{Yasuda2020}%
  \BibitemOpen
  \bibfield  {author} {\bibinfo {author} {\bibfnamefont {K.}~\bibnamefont
  {Yasuda}}, \bibinfo {author} {\bibfnamefont {T.}~\bibnamefont {Morimoto}},
  \bibinfo {author} {\bibfnamefont {R.}~\bibnamefont {Yoshimi}}, \bibinfo
  {author} {\bibfnamefont {M.}~\bibnamefont {Mogi}}, \bibinfo {author}
  {\bibfnamefont {A.}~\bibnamefont {Tsukazaki}}, \bibinfo {author}
  {\bibfnamefont {M.}~\bibnamefont {Kawamura}}, \bibinfo {author}
  {\bibfnamefont {K.~S.}\ \bibnamefont {Takahashi}}, \bibinfo {author}
  {\bibfnamefont {M.}~\bibnamefont {Kawasaki}}, \bibinfo {author}
  {\bibfnamefont {N.}~\bibnamefont {Nagaosa}},\ and\ \bibinfo {author}
  {\bibfnamefont {Y.}~\bibnamefont {Tokura}},\ }\bibfield  {title} {\bibinfo
  {title} \textit{Large non-reciprocal charge transport mediated by quantum anomalous
  {Hall} edge states},\ }\href@noop {} {\bibfield  {journal} {\bibinfo
  {journal} {Nat. Nanotechnology}\ }\textbf {\bibinfo {volume} {15}},\ \bibinfo
  {pages} {831} (\bibinfo {year} {2020})}\BibitemShut {NoStop}%
\bibitem [{\citenamefont {Koster}\ \emph {et~al.}(2012)\citenamefont {Koster},
  \citenamefont {Klein}, \citenamefont {Siemons}, \citenamefont {Rijnders},
  \citenamefont {Dodge}, \citenamefont {Eom}, \citenamefont {Blank},\ and\
  \citenamefont {Beasley}}]{Koster2012}%
  \BibitemOpen
  \bibfield  {author} {\bibinfo {author} {\bibfnamefont {G.}~\bibnamefont
  {Koster}}, \bibinfo {author} {\bibfnamefont {L.}~\bibnamefont {Klein}},
  \bibinfo {author} {\bibfnamefont {W.}~\bibnamefont {Siemons}}, \bibinfo
  {author} {\bibfnamefont {G.}~\bibnamefont {Rijnders}}, \bibinfo {author}
  {\bibfnamefont {J.~S.}\ \bibnamefont {Dodge}}, \bibinfo {author}
  {\bibfnamefont {C.-B.}\ \bibnamefont {Eom}}, \bibinfo {author} {\bibfnamefont
  {D.~H.~A.}\ \bibnamefont {Blank}},\ and\ \bibinfo {author} {\bibfnamefont
  {M.~R.}\ \bibnamefont {Beasley}},\ }\bibfield  {title} {\bibinfo {title}
  \textit{Structure, physical properties, and applications of {SrRuO$_{3}$} thin
  films},\ }\href@noop {} {\bibfield  {journal} {\bibinfo  {journal} {Rev. Mod.
  Phys.}\ }\textbf {\bibinfo {volume} {84}},\ \bibinfo {pages} {253} (\bibinfo
  {year} {2012})}\BibitemShut {NoStop}%
\bibitem [{\citenamefont {Kar}\ \emph {et~al.}(2021)\citenamefont {Kar},
  \citenamefont {Singh}, \citenamefont {Yang}, \citenamefont {Lin},
  \citenamefont {Das}, \citenamefont {Hsu},\ and\ \citenamefont
  {Lee}}]{Kar2021}%
  \BibitemOpen
  \bibfield  {author} {\bibinfo {author} {\bibfnamefont {U.}~\bibnamefont
  {Kar}}, \bibinfo {author} {\bibfnamefont {A.~K.}\ \bibnamefont {Singh}},
  \bibinfo {author} {\bibfnamefont {S.}~\bibnamefont {Yang}}, \bibinfo {author}
  {\bibfnamefont {C.-Y.}\ \bibnamefont {Lin}}, \bibinfo {author} {\bibfnamefont
  {B.}~\bibnamefont {Das}}, \bibinfo {author} {\bibfnamefont {C.-H.}\
  \bibnamefont {Hsu}},\ and\ \bibinfo {author} {\bibfnamefont {W.-L.}\
  \bibnamefont {Lee}},\ }\bibfield  {title} {\bibinfo {title} \textit{High-sensitivity
  of initial {SrO} growth on the residual resistivity in epitaxial thin films
  of {SrRuO$_3$} on {SrTiO$_3$ (001)}},\ }\href@noop {} {\bibfield  {journal}
  {\bibinfo  {journal} {Sci. Rep.}\ }\textbf {\bibinfo {volume} {11}},\
  \bibinfo {pages} {16070} (\bibinfo {year} {2021})}\BibitemShut {NoStop}%
\bibitem [{\citenamefont {Fang}\ \emph {et~al.}(2003)\citenamefont {Fang},
  \citenamefont {Nagaosa}, \citenamefont {Takahashi}, \citenamefont {Asamitsu},
  \citenamefont {Mathieu}, \citenamefont {Ogasawara}, \citenamefont {Yamada},
  \citenamefont {Kawasaki}, \citenamefont {Tokura},\ and\ \citenamefont
  {Terakura}}]{Fang2003}%
  \BibitemOpen
  \bibfield  {author} {\bibinfo {author} {\bibfnamefont {Z.}~\bibnamefont
  {Fang}}, \bibinfo {author} {\bibfnamefont {N.}~\bibnamefont {Nagaosa}},
  \bibinfo {author} {\bibfnamefont {K.~S.}\ \bibnamefont {Takahashi}}, \bibinfo
  {author} {\bibfnamefont {A.}~\bibnamefont {Asamitsu}}, \bibinfo {author}
  {\bibfnamefont {R.}~\bibnamefont {Mathieu}}, \bibinfo {author} {\bibfnamefont
  {T.}~\bibnamefont {Ogasawara}}, \bibinfo {author} {\bibfnamefont
  {H.}~\bibnamefont {Yamada}}, \bibinfo {author} {\bibfnamefont
  {M.}~\bibnamefont {Kawasaki}}, \bibinfo {author} {\bibfnamefont
  {Y.}~\bibnamefont {Tokura}},\ and\ \bibinfo {author} {\bibfnamefont
  {K.}~\bibnamefont {Terakura}},\ }\bibfield  {title} {\bibinfo {title} \textit{The
  anomalous {Hall} effect and magnetic monopoles in momentum space},\
  }\href@noop {} {\bibfield  {journal} {\bibinfo  {journal} {Science}\ }\textbf
  {\bibinfo {volume} {302}},\ \bibinfo {pages} {92} (\bibinfo {year}
  {2003})}\BibitemShut {NoStop}%
\bibitem [{\citenamefont {Chen}\ \emph {et~al.}(2013)\citenamefont {Chen},
  \citenamefont {Bergman},\ and\ \citenamefont {Burkov}}]{Chen2013}%
  \BibitemOpen
  \bibfield  {author} {\bibinfo {author} {\bibfnamefont {Y.}~\bibnamefont
  {Chen}}, \bibinfo {author} {\bibfnamefont {D.~L.}\ \bibnamefont {Bergman}},\
  and\ \bibinfo {author} {\bibfnamefont {A.~A.}\ \bibnamefont {Burkov}},\
  }\bibfield  {title} {\bibinfo {title} \textit{Weyl fermions and the anomalous {Hall}
  effect in metallic ferromagnets},\ }\href@noop {} {\bibfield  {journal}
  {\bibinfo  {journal} {Phys. Rev. B}\ }\textbf {\bibinfo {volume} {88}},\
  \bibinfo {pages} {125110} (\bibinfo {year} {2013})}\BibitemShut {NoStop}%
\bibitem [{\citenamefont {Itoh}\ \emph {et~al.}(2016)\citenamefont {Itoh},
  \citenamefont {Endoh}, \citenamefont {Yokoo}, \citenamefont {Ibuka},
  \citenamefont {Park}, \citenamefont {Kaneko}, \citenamefont {Takahashi},
  \citenamefont {Tokura},\ and\ \citenamefont {Nagaosa}}]{Itoh2016}%
  \BibitemOpen
  \bibfield  {author} {\bibinfo {author} {\bibfnamefont {S.}~\bibnamefont
  {Itoh}}, \bibinfo {author} {\bibfnamefont {Y.}~\bibnamefont {Endoh}},
  \bibinfo {author} {\bibfnamefont {T.}~\bibnamefont {Yokoo}}, \bibinfo
  {author} {\bibfnamefont {S.}~\bibnamefont {Ibuka}}, \bibinfo {author}
  {\bibfnamefont {J.-G.}\ \bibnamefont {Park}}, \bibinfo {author}
  {\bibfnamefont {Y.}~\bibnamefont {Kaneko}}, \bibinfo {author} {\bibfnamefont
  {K.~S.}\ \bibnamefont {Takahashi}}, \bibinfo {author} {\bibfnamefont
  {Y.}~\bibnamefont {Tokura}},\ and\ \bibinfo {author} {\bibfnamefont
  {N.}~\bibnamefont {Nagaosa}},\ }\bibfield  {title} {\bibinfo {title} \textit{Weyl
  fermions and spin dynamics of metallic ferromagnet {SrRuO$_3$}},\ }\href@noop
  {} {\bibfield  {journal} {\bibinfo  {journal} {Nat. Commun.}\ }\textbf
  {\bibinfo {volume} {7}},\ \bibinfo {pages} {11788} (\bibinfo {year}
  {2016})}\BibitemShut {NoStop}%
\bibitem [{\citenamefont {Jenni}\ \emph {et~al.}(2019)\citenamefont {Jenni},
  \citenamefont {Kunkem\"oller}, \citenamefont {Br\"uning}, \citenamefont
  {Lorenz}, \citenamefont {Sidis}, \citenamefont {Schneidewind}, \citenamefont
  {Nugroho}, \citenamefont {Rosch}, \citenamefont {Khomskii},\ and\
  \citenamefont {Braden}}]{Jenni2019}%
  \BibitemOpen
  \bibfield  {author} {\bibinfo {author} {\bibfnamefont {K.}~\bibnamefont
  {Jenni}}, \bibinfo {author} {\bibfnamefont {S.}~\bibnamefont
  {Kunkem\"oller}}, \bibinfo {author} {\bibfnamefont {D.}~\bibnamefont
  {Br\"uning}}, \bibinfo {author} {\bibfnamefont {T.}~\bibnamefont {Lorenz}},
  \bibinfo {author} {\bibfnamefont {Y.}~\bibnamefont {Sidis}}, \bibinfo
  {author} {\bibfnamefont {A.}~\bibnamefont {Schneidewind}}, \bibinfo {author}
  {\bibfnamefont {A.~A.}\ \bibnamefont {Nugroho}}, \bibinfo {author}
  {\bibfnamefont {A.}~\bibnamefont {Rosch}}, \bibinfo {author} {\bibfnamefont
  {D.~I.}\ \bibnamefont {Khomskii}},\ and\ \bibinfo {author} {\bibfnamefont
  {M.}~\bibnamefont {Braden}},\ }\bibfield  {title} {\bibinfo {title}
  \textit{Interplay of electronic and spin degrees in ferromagnetic {SrRuO$_{3}$}:
  Anomalous softening of the magnon gap and stiffness},\ }\href@noop {}
  {\bibfield  {journal} {\bibinfo  {journal} {Phys. Rev. Lett.}\ }\textbf
  {\bibinfo {volume} {123}},\ \bibinfo {pages} {017202} (\bibinfo {year}
  {2019})}\BibitemShut {NoStop}%
\bibitem [{\citenamefont {Nair}\ \emph {et~al.}(2018)\citenamefont {Nair},
  \citenamefont {Liu}, \citenamefont {Ruf}, \citenamefont {Schreiber},
  \citenamefont {Shang}, \citenamefont {Baek}, \citenamefont {Goodge},
  \citenamefont {Kourkoutis}, \citenamefont {Liu}, \citenamefont {Shen},\ and\
  \citenamefont {Schlom}}]{Nair2018}%
  \BibitemOpen
  \bibfield  {author} {\bibinfo {author} {\bibfnamefont {H.~P.}\ \bibnamefont
  {Nair}}, \bibinfo {author} {\bibfnamefont {Y.}~\bibnamefont {Liu}}, \bibinfo
  {author} {\bibfnamefont {J.~P.}\ \bibnamefont {Ruf}}, \bibinfo {author}
  {\bibfnamefont {N.~J.}\ \bibnamefont {Schreiber}}, \bibinfo {author}
  {\bibfnamefont {S.-L.}\ \bibnamefont {Shang}}, \bibinfo {author}
  {\bibfnamefont {D.~J.}\ \bibnamefont {Baek}}, \bibinfo {author}
  {\bibfnamefont {B.~H.}\ \bibnamefont {Goodge}}, \bibinfo {author}
  {\bibfnamefont {L.~F.}\ \bibnamefont {Kourkoutis}}, \bibinfo {author}
  {\bibfnamefont {Z.-K.}\ \bibnamefont {Liu}}, \bibinfo {author} {\bibfnamefont
  {K.~M.}\ \bibnamefont {Shen}},\ and\ \bibinfo {author} {\bibfnamefont
  {D.~G.}\ \bibnamefont {Schlom}},\ }\bibfield  {title} {\bibinfo {title}
  \textit{Synthesis science of {SrRuO$_3$} and {CaRuO$_3$} epitaxial films with high
  residual resistivity ratios},\ }\href@noop {} {\bibfield  {journal} {\bibinfo
   {journal} {APL Mater.}\ }\textbf {\bibinfo {volume} {6}},\ \bibinfo {pages}
  {046101} (\bibinfo {year} {2018})}\BibitemShut {NoStop}%
\bibitem [{\citenamefont {Takiguchi}\ \emph {et~al.}(2020)\citenamefont
  {Takiguchi}, \citenamefont {Wakabayashi}, \citenamefont {Irie}, \citenamefont
  {Krockenberger}, \citenamefont {Otsuka}, \citenamefont {Sawada},
  \citenamefont {Nikolaev}, \citenamefont {Das}, \citenamefont {Tanaka},
  \citenamefont {Taniyasu},\ and\ \citenamefont {Yamamoto}}]{Taki2020}%
  \BibitemOpen
  \bibfield  {author} {\bibinfo {author} {\bibfnamefont {K.}~\bibnamefont
  {Takiguchi}}, \bibinfo {author} {\bibfnamefont {Y.~K.}\ \bibnamefont
  {Wakabayashi}}, \bibinfo {author} {\bibfnamefont {H.}~\bibnamefont {Irie}},
  \bibinfo {author} {\bibfnamefont {Y.}~\bibnamefont {Krockenberger}}, \bibinfo
  {author} {\bibfnamefont {T.}~\bibnamefont {Otsuka}}, \bibinfo {author}
  {\bibfnamefont {H.}~\bibnamefont {Sawada}}, \bibinfo {author} {\bibfnamefont
  {S.~A.}\ \bibnamefont {Nikolaev}}, \bibinfo {author} {\bibfnamefont
  {H.}~\bibnamefont {Das}}, \bibinfo {author} {\bibfnamefont {M.}~\bibnamefont
  {Tanaka}}, \bibinfo {author} {\bibfnamefont {Y.}~\bibnamefont {Taniyasu}},\
  and\ \bibinfo {author} {\bibfnamefont {H.}~\bibnamefont {Yamamoto}},\
  }\bibfield  {title} {\bibinfo {title} \textit{Quantum transport evidence of {Weyl}
  fermions in an epitaxial ferromagnetic oxide},\ }\href@noop {} {\bibfield
  {journal} {\bibinfo  {journal} {Nat. Commun.}\ }\textbf {\bibinfo {volume}
  {11}},\ \bibinfo {pages} {4969} (\bibinfo {year} {2020})}\BibitemShut
  {NoStop}%
\bibitem [{\citenamefont {Capogna}\ \emph {et~al.}(2002)\citenamefont
  {Capogna}, \citenamefont {Mackenzie}, \citenamefont {Perry}, \citenamefont
  {Grigera}, \citenamefont {Galvin}, \citenamefont {Raychaudhuri},
  \citenamefont {Schofield}, \citenamefont {Alexander}, \citenamefont {Cao},
  \citenamefont {Julian},\ and\ \citenamefont {Maeno}}]{Cap2002}%
  \BibitemOpen
  \bibfield  {author} {\bibinfo {author} {\bibfnamefont {L.}~\bibnamefont
  {Capogna}}, \bibinfo {author} {\bibfnamefont {A.~P.}\ \bibnamefont
  {Mackenzie}}, \bibinfo {author} {\bibfnamefont {R.~S.}\ \bibnamefont
  {Perry}}, \bibinfo {author} {\bibfnamefont {S.~A.}\ \bibnamefont {Grigera}},
  \bibinfo {author} {\bibfnamefont {L.~M.}\ \bibnamefont {Galvin}}, \bibinfo
  {author} {\bibfnamefont {P.}~\bibnamefont {Raychaudhuri}}, \bibinfo {author}
  {\bibfnamefont {A.~J.}\ \bibnamefont {Schofield}}, \bibinfo {author}
  {\bibfnamefont {C.~S.}\ \bibnamefont {Alexander}}, \bibinfo {author}
  {\bibfnamefont {G.}~\bibnamefont {Cao}}, \bibinfo {author} {\bibfnamefont
  {S.~R.}\ \bibnamefont {Julian}},\ and\ \bibinfo {author} {\bibfnamefont
  {Y.}~\bibnamefont {Maeno}},\ }\bibfield  {title} {\bibinfo {title}
  \textit{Sensitivity to disorder of the metallic state in the ruthenates},\
  }\href@noop {} {\bibfield  {journal} {\bibinfo  {journal} {Phys. Rev. Lett.}\
  }\textbf {\bibinfo {volume} {88}},\ \bibinfo {pages} {076602} (\bibinfo
  {year} {2002})}\BibitemShut {NoStop}%
\bibitem [{\citenamefont {Nandkishore}\ \emph {et~al.}(2014)\citenamefont
  {Nandkishore}, \citenamefont {Huse},\ and\ \citenamefont
  {Sondhi}}]{Nand2014}%
  \BibitemOpen
  \bibfield  {author} {\bibinfo {author} {\bibfnamefont {R.}~\bibnamefont
  {Nandkishore}}, \bibinfo {author} {\bibfnamefont {D.~A.}\ \bibnamefont
  {Huse}},\ and\ \bibinfo {author} {\bibfnamefont {S.~L.}\ \bibnamefont
  {Sondhi}},\ }\bibfield  {title} {\bibinfo {title} \textit{Rare region effects
  dominate weakly disordered three-dimensional {Dirac} points},\ }\href@noop {}
  {\bibfield  {journal} {\bibinfo  {journal} {Phys. Rev. B}\ }\textbf {\bibinfo
  {volume} {89}},\ \bibinfo {pages} {245110} (\bibinfo {year}
  {2014})}\BibitemShut {NoStop}%
\bibitem [{\citenamefont {Kaneta-Takada}\ \emph {et~al.}(2022)\citenamefont
  {Kaneta-Takada}, \citenamefont {Wakabayashi}, \citenamefont {Krockenberger},
  \citenamefont {Nomura}, \citenamefont {Kohama}, \citenamefont {Nikolaev},
  \citenamefont {Das}, \citenamefont {Irie}, \citenamefont {Takiguchi},
  \citenamefont {Ohya}, \citenamefont {Tanaka}, \citenamefont {Taniyasu},\ and\
  \citenamefont {Yamamoto}}]{Kaneta2022}%
  \BibitemOpen
  \bibfield  {author} {\bibinfo {author} {\bibfnamefont {S.}~\bibnamefont
  {Kaneta-Takada}}, \bibinfo {author} {\bibfnamefont {Y.~K.}\ \bibnamefont
  {Wakabayashi}}, \bibinfo {author} {\bibfnamefont {Y.}~\bibnamefont
  {Krockenberger}}, \bibinfo {author} {\bibfnamefont {T.}~\bibnamefont
  {Nomura}}, \bibinfo {author} {\bibfnamefont {Y.}~\bibnamefont {Kohama}},
  \bibinfo {author} {\bibfnamefont {S.~A.}\ \bibnamefont {Nikolaev}}, \bibinfo
  {author} {\bibfnamefont {H.}~\bibnamefont {Das}}, \bibinfo {author}
  {\bibfnamefont {H.}~\bibnamefont {Irie}}, \bibinfo {author} {\bibfnamefont
  {K.}~\bibnamefont {Takiguchi}}, \bibinfo {author} {\bibfnamefont
  {S.}~\bibnamefont {Ohya}}, \bibinfo {author} {\bibfnamefont {M.}~\bibnamefont
  {Tanaka}}, \bibinfo {author} {\bibfnamefont {Y.}~\bibnamefont {Taniyasu}},\
  and\ \bibinfo {author} {\bibfnamefont {H.}~\bibnamefont {Yamamoto}},\
  }\bibfield  {title} {\bibinfo {title} \textit{High-mobility two-dimensional carriers
  from surface {Fermi} arcs in magnetic {Weyl} semimetal films},\ }\href@noop
  {} {\bibfield  {journal} {\bibinfo  {journal} {npj Quantum Mater.}\ }\textbf
  {\bibinfo {volume} {7}},\ \bibinfo {pages} {102} (\bibinfo {year}
  {2022})}\BibitemShut {NoStop}%
\bibitem [{\citenamefont {Kar}\ \emph {et~al.}(2023)\citenamefont {Kar},
  \citenamefont {Singh}, \citenamefont {Hsu}, \citenamefont {Lin},
  \citenamefont {Das}, \citenamefont {Cheng}, \citenamefont {Berben},
  \citenamefont {Yang}, \citenamefont {Lin}, \citenamefont {Hsu}, \citenamefont
  {Wiedmann}, \citenamefont {Lee},\ and\ \citenamefont {Lee}}]{kar2022}%
  \BibitemOpen
  \bibfield  {author} {\bibinfo {author} {\bibfnamefont {U.}~\bibnamefont
  {Kar}}, \bibinfo {author} {\bibfnamefont {A.~K.}\ \bibnamefont {Singh}},
  \bibinfo {author} {\bibfnamefont {Y.-T.}\ \bibnamefont {Hsu}}, \bibinfo
  {author} {\bibfnamefont {C.-Y.}\ \bibnamefont {Lin}}, \bibinfo {author}
  {\bibfnamefont {B.}~\bibnamefont {Das}}, \bibinfo {author} {\bibfnamefont
  {C.-T.}\ \bibnamefont {Cheng}}, \bibinfo {author} {\bibfnamefont
  {M.}~\bibnamefont {Berben}}, \bibinfo {author} {\bibfnamefont
  {S.}~\bibnamefont {Yang}}, \bibinfo {author} {\bibfnamefont {C.-Y.}\
  \bibnamefont {Lin}}, \bibinfo {author} {\bibfnamefont {C.-H.}\ \bibnamefont
  {Hsu}}, \bibinfo {author} {\bibfnamefont {S.}~\bibnamefont {Wiedmann}},
  \bibinfo {author} {\bibfnamefont {W.-C.}\ \bibnamefont {Lee}},\ and\ \bibinfo
  {author} {\bibfnamefont {W.-L.}\ \bibnamefont {Lee}},\ }\bibfield  {title}
  {\bibinfo {title} \textit{The thickness dependence of quantum oscillations in
  ferromagnetic {Weyl} metal {SrRuO$_{3}$}},\ }\href@noop {} {\bibfield
  {journal} {\bibinfo  {journal} {npj Quantum Mater.}\ }\textbf {\bibinfo
  {volume} {8}},\ \bibinfo {pages} {8} (\bibinfo {year} {2023})}\BibitemShut
  {NoStop}%
\bibitem{SOM} See Supplemental Material at http://link.aps.org/supplemental/10.1103/PhysRevX.14.011022 for thin film characterizations, device geometry descriptions, electronic and magnetic structure calculations as well as Berry curvature dipole and quantum metric dipole calculations.
\bibitem [{\citenamefont {Nagaosa}\ \emph {et~al.}(2010)\citenamefont
  {Nagaosa}, \citenamefont {Sinova}, \citenamefont {Onoda}, \citenamefont
  {MacDonald},\ and\ \citenamefont {Ong}}]{Nagaosa2010}%
  \BibitemOpen
  \bibfield  {author} {\bibinfo {author} {\bibfnamefont {N.}~\bibnamefont
  {Nagaosa}}, \bibinfo {author} {\bibfnamefont {J.}~\bibnamefont {Sinova}},
  \bibinfo {author} {\bibfnamefont {S.}~\bibnamefont {Onoda}}, \bibinfo
  {author} {\bibfnamefont {A.~H.}\ \bibnamefont {MacDonald}},\ and\ \bibinfo
  {author} {\bibfnamefont {N.~P.}\ \bibnamefont {Ong}},\ }\bibfield  {title}
  {\bibinfo {title} \textit{Anomalous {Hall} effect},\ }\href@noop {} {\bibfield
  {journal} {\bibinfo  {journal} {Rev. Mod. Phys.}\ }\textbf {\bibinfo {volume}
  {82}},\ \bibinfo {pages} {1539} (\bibinfo {year} {2010})}\BibitemShut
  {NoStop}%
\bibitem [{\citenamefont {Raoux}\ \emph {et~al.}(2014)\citenamefont {Raoux},
  \citenamefont {Morigi}, \citenamefont {Fuchs}, \citenamefont {Pi\'echon},\
  and\ \citenamefont {Montambaux}}]{Rao2014}%
  \BibitemOpen
  \bibfield  {author} {\bibinfo {author} {\bibfnamefont {A.}~\bibnamefont
  {Raoux}}, \bibinfo {author} {\bibfnamefont {M.}~\bibnamefont {Morigi}},
  \bibinfo {author} {\bibfnamefont {J.-N.}\ \bibnamefont {Fuchs}}, \bibinfo
  {author} {\bibfnamefont {F.}~\bibnamefont {Pi\'echon}},\ and\ \bibinfo
  {author} {\bibfnamefont {G.}~\bibnamefont {Montambaux}},\ }\bibfield  {title}
  {\bibinfo {title} \textit{From dia- to paramagnetic orbital susceptibility of
  massless fermions},\ }\href@noop {} {\bibfield  {journal} {\bibinfo
  {journal} {Phys. Rev. Lett.}\ }\textbf {\bibinfo {volume} {112}},\ \bibinfo
  {pages} {026402} (\bibinfo {year} {2014})}\BibitemShut {NoStop}%
\bibitem [{\citenamefont {Suetsugu}\ \emph {et~al.}(2021)\citenamefont
  {Suetsugu}, \citenamefont {Kitagawa}, \citenamefont {Kariyado}, \citenamefont
  {Rost}, \citenamefont {Nuss}, \citenamefont {M\"uhle}, \citenamefont
  {Ogata},\ and\ \citenamefont {Takagi}}]{Sue2021}%
  \BibitemOpen
  \bibfield  {author} {\bibinfo {author} {\bibfnamefont {S.}~\bibnamefont
  {Suetsugu}}, \bibinfo {author} {\bibfnamefont {K.}~\bibnamefont {Kitagawa}},
  \bibinfo {author} {\bibfnamefont {T.}~\bibnamefont {Kariyado}}, \bibinfo
  {author} {\bibfnamefont {A.~W.}\ \bibnamefont {Rost}}, \bibinfo {author}
  {\bibfnamefont {J.}~\bibnamefont {Nuss}}, \bibinfo {author} {\bibfnamefont
  {C.}~\bibnamefont {M\"uhle}}, \bibinfo {author} {\bibfnamefont
  {M.}~\bibnamefont {Ogata}},\ and\ \bibinfo {author} {\bibfnamefont
  {H.}~\bibnamefont {Takagi}},\ }\bibfield  {title} {\bibinfo {title} \textit{Giant
  orbital diamagnetism of three-dimensional {Dirac} electrons in
  {Sr$_{3}${PbO}} antiperovskite},\ }\href@noop {} {\bibfield  {journal}
  {\bibinfo  {journal} {Phys. Rev. B}\ }\textbf {\bibinfo {volume} {103}},\
  \bibinfo {pages} {115117} (\bibinfo {year} {2021})}\BibitemShut {NoStop}%
\bibitem [{\citenamefont {Morimoto}\ and\ \citenamefont
  {Nagaosa}(2016)}]{Morimoto2016}%
  \BibitemOpen
  \bibfield  {author} {\bibinfo {author} {\bibfnamefont {T.}~\bibnamefont
  {Morimoto}}\ and\ \bibinfo {author} {\bibfnamefont {N.}~\bibnamefont
  {Nagaosa}},\ }\bibfield  {title} {\bibinfo {title} \textit{Chiral anomaly and giant
  magnetochiral anisotropy in noncentrosymmetric {Weyl} semimetals},\
  }\href@noop {} {\bibfield  {journal} {\bibinfo  {journal} {Phys. Rev. Lett.}\
  }\textbf {\bibinfo {volume} {117}},\ \bibinfo {pages} {146603} (\bibinfo
  {year} {2016})}\BibitemShut {NoStop}%
\bibitem [{\citenamefont {Li}\ \emph {et~al.}(2021)\citenamefont {Li},
  \citenamefont {Heinonen}, \citenamefont {Burkov},\ and\ \citenamefont
  {Zhang}}]{LiRH2021}%
  \BibitemOpen
  \bibfield  {author} {\bibinfo {author} {\bibfnamefont {R.-H.}\ \bibnamefont
  {Li}}, \bibinfo {author} {\bibfnamefont {O.~G.}\ \bibnamefont {Heinonen}},
  \bibinfo {author} {\bibfnamefont {A.~A.}\ \bibnamefont {Burkov}},\ and\
  \bibinfo {author} {\bibfnamefont {S.~S.-L.}\ \bibnamefont {Zhang}},\
  }\bibfield  {title} {\bibinfo {title} \textit{Nonlinear {Hall} effect in {Weyl}
  semimetals induced by chiral anomaly},\ }\href@noop {} {\bibfield  {journal}
  {\bibinfo  {journal} {Phys. Rev. B}\ }\textbf {\bibinfo {volume} {103}},\
  \bibinfo {pages} {045105} (\bibinfo {year} {2021})}\BibitemShut {NoStop}%
\bibitem [{\citenamefont {Nandy}\ \emph {et~al.}(2021)\citenamefont {Nandy},
  \citenamefont {Zeng},\ and\ \citenamefont {Tewari}}]{Nandy2021}%
  \BibitemOpen
  \bibfield  {author} {\bibinfo {author} {\bibfnamefont {S.}~\bibnamefont
  {Nandy}}, \bibinfo {author} {\bibfnamefont {C.}~\bibnamefont {Zeng}},\ and\
  \bibinfo {author} {\bibfnamefont {S.}~\bibnamefont {Tewari}},\ }\bibfield
  {title} {\bibinfo {title} \textit{Chiral anomaly induced nonlinear {Hall} effect in
  semimetals with multiple {Weyl} points},\ }\href@noop {} {\bibfield
  {journal} {\bibinfo  {journal} {Phys. Rev. B}\ }\textbf {\bibinfo {volume}
  {104}},\ \bibinfo {pages} {205124} (\bibinfo {year} {2021})}\BibitemShut
  {NoStop}%
\bibitem [{\citenamefont {Du}\ \emph {et~al.}(2019)\citenamefont {Du},
  \citenamefont {Wang}, \citenamefont {Li}, \citenamefont {Lu},\ and\
  \citenamefont {Xie}}]{Du2019}%
  \BibitemOpen
  \bibfield  {author} {\bibinfo {author} {\bibfnamefont {Z.~Z.}\ \bibnamefont
  {Du}}, \bibinfo {author} {\bibfnamefont {C.~M.}\ \bibnamefont {Wang}},
  \bibinfo {author} {\bibfnamefont {S.}~\bibnamefont {Li}}, \bibinfo {author}
  {\bibfnamefont {H.-Z.}\ \bibnamefont {Lu}},\ and\ \bibinfo {author}
  {\bibfnamefont {X.~C.}\ \bibnamefont {Xie}},\ }\bibfield  {title} {\bibinfo
  {title} \textit{Disorder-induced nonlinear {Hall} effect with time-reversal
  symmetry},\ }\href@noop {} {\bibfield  {journal} {\bibinfo  {journal} {Nat.
  Commun.}\ }\textbf {\bibinfo {volume} {10}},\ \bibinfo {pages} {3047}
  (\bibinfo {year} {2019})}\BibitemShut {NoStop}%
\bibitem [{\citenamefont {Isobe}\ \emph {et~al.}(2020)\citenamefont {Isobe},
  \citenamefont {Xu},\ and\ \citenamefont {Fu}}]{Iso2020}%
  \BibitemOpen
  \bibfield  {author} {\bibinfo {author} {\bibfnamefont {H.}~\bibnamefont
  {Isobe}}, \bibinfo {author} {\bibfnamefont {S.-Y.}\ \bibnamefont {Xu}},\ and\
  \bibinfo {author} {\bibfnamefont {L.}~\bibnamefont {Fu}},\ }\bibfield
  {title} {\bibinfo {title} \textit{High-frequency rectification via chiral {Bloch}
  electrons},\ }\href@noop {} {\bibfield  {journal} {\bibinfo  {journal} {Sci.
  Adv.}\ }\textbf {\bibinfo {volume} {6}},\ \bibinfo {pages} {eaay2497}
  (\bibinfo {year} {2020})}\BibitemShut {NoStop}%
\bibitem [{\citenamefont {He}\ \emph {et~al.}(2021)\citenamefont {He},
  \citenamefont {Isobe}, \citenamefont {Zhu}, \citenamefont {Hsu},
  \citenamefont {Fu},\ and\ \citenamefont {Yang}}]{He2021}%
  \BibitemOpen
  \bibfield  {author} {\bibinfo {author} {\bibfnamefont {P.}~\bibnamefont
  {He}}, \bibinfo {author} {\bibfnamefont {H.}~\bibnamefont {Isobe}}, \bibinfo
  {author} {\bibfnamefont {D.}~\bibnamefont {Zhu}}, \bibinfo {author}
  {\bibfnamefont {C.-H.}\ \bibnamefont {Hsu}}, \bibinfo {author} {\bibfnamefont
  {L.}~\bibnamefont {Fu}},\ and\ \bibinfo {author} {\bibfnamefont
  {H.}~\bibnamefont {Yang}},\ }\bibfield  {title} {\bibinfo {title} \textit{Quantum
  frequency doubling in the topological insulator {Bi$_2$Se$_3$}},\ }\href@noop
  {} {\bibfield  {journal} {\bibinfo  {journal} {Nat. Commun.}\ }\textbf
  {\bibinfo {volume} {12}},\ \bibinfo {pages} {698} (\bibinfo {year}
  {2021})}\BibitemShut {NoStop}%
\bibitem [{\citenamefont {Wang}\ \emph {et~al.}(2021)\citenamefont {Wang},
  \citenamefont {Gao},\ and\ \citenamefont {Xiao}}]{Wang2021}%
  \BibitemOpen
  \bibfield  {author} {\bibinfo {author} {\bibfnamefont {C.}~\bibnamefont
  {Wang}}, \bibinfo {author} {\bibfnamefont {Y.}~\bibnamefont {Gao}},\ and\
  \bibinfo {author} {\bibfnamefont {D.}~\bibnamefont {Xiao}},\ }\bibfield
  {title} {\bibinfo {title} \textit{Intrinsic nonlinear {Hall} effect in
  antiferromagnetic tetragonal {CuMnAs}},\ }\href@noop {} {\bibfield  {journal}
  {\bibinfo  {journal} {Phys. Rev. Lett.}\ }\textbf {\bibinfo {volume} {127}},\
  \bibinfo {pages} {277201} (\bibinfo {year} {2021})}\BibitemShut {NoStop}%
\bibitem [{\citenamefont {Liu}\ \emph {et~al.}(2021)\citenamefont {Liu},
  \citenamefont {Zhao}, \citenamefont {Huang}, \citenamefont {Wu},
  \citenamefont {Sheng}, \citenamefont {Xiao},\ and\ \citenamefont
  {Yang}}]{Liu2021}%
  \BibitemOpen
  \bibfield  {author} {\bibinfo {author} {\bibfnamefont {H.}~\bibnamefont
  {Liu}}, \bibinfo {author} {\bibfnamefont {J.}~\bibnamefont {Zhao}}, \bibinfo
  {author} {\bibfnamefont {Y.-X.}\ \bibnamefont {Huang}}, \bibinfo {author}
  {\bibfnamefont {W.}~\bibnamefont {Wu}}, \bibinfo {author} {\bibfnamefont
  {X.-L.}\ \bibnamefont {Sheng}}, \bibinfo {author} {\bibfnamefont
  {C.}~\bibnamefont {Xiao}},\ and\ \bibinfo {author} {\bibfnamefont {S.~A.}\
  \bibnamefont {Yang}},\ }\bibfield  {title} {\bibinfo {title} \textit{Intrinsic
  second-order anomalous {Hall} effect and its application in compensated
  antiferromagnets},\ }\href@noop {} {\bibfield  {journal} {\bibinfo  {journal}
  {Phys. Rev. Lett.}\ }\textbf {\bibinfo {volume} {127}},\ \bibinfo {pages}
  {277202} (\bibinfo {year} {2021})}\BibitemShut {NoStop}%
\bibitem [{\citenamefont {Gao}\ \emph {et~al.}(2023)\citenamefont {Gao},
  \citenamefont {Liu}, \citenamefont {Qiu}, \citenamefont {Ghosh},
  \citenamefont {Trevisan}, \citenamefont {Onishi}, \citenamefont {Hu},
  \citenamefont {Qian}, \citenamefont {Tien}, \citenamefont {Chen},
  \citenamefont {Huang}, \citenamefont {Bérubé}, \citenamefont {Li},
  \citenamefont {Tzschaschel}, \citenamefont {Dinh}, \citenamefont {Sun},
  \citenamefont {Ho}, \citenamefont {Lien}, \citenamefont {Singh},
  \citenamefont {Watanabe}, \citenamefont {Taniguchi}, \citenamefont {Bell},
  \citenamefont {Lin}, \citenamefont {Chang}, \citenamefont {Du}, \citenamefont
  {Bansil}, \citenamefont {Fu}, \citenamefont {Ni}, \citenamefont {Orth},
  \citenamefont {Ma},\ and\ \citenamefont {Xu}}]{Gao2023}%
  \BibitemOpen
  \bibfield  {author} {\bibinfo {author} {\bibfnamefont {A.}~\bibnamefont
  {Gao}}, \bibinfo {author} {\bibfnamefont {Y.-F.}\ \bibnamefont {Liu}},
  \bibinfo {author} {\bibfnamefont {J.-X.}\ \bibnamefont {Qiu}}, \bibinfo
  {author} {\bibfnamefont {B.}~\bibnamefont {Ghosh}}, \bibinfo {author}
  {\bibfnamefont {T.~V.}\ \bibnamefont {Trevisan}}, \bibinfo {author}
  {\bibfnamefont {Y.}~\bibnamefont {Onishi}}, \bibinfo {author} {\bibfnamefont
  {C.}~\bibnamefont {Hu}}, \bibinfo {author} {\bibfnamefont {T.}~\bibnamefont
  {Qian}}, \bibinfo {author} {\bibfnamefont {H.-J.}\ \bibnamefont {Tien}},
  \bibinfo {author} {\bibfnamefont {S.-W.}\ \bibnamefont {Chen}}, \bibinfo
  {author} {\bibfnamefont {M.}~\bibnamefont {Huang}}, \bibinfo {author}
  {\bibfnamefont {D.}~\bibnamefont {Bérubé}}, \bibinfo {author}
  {\bibfnamefont {H.}~\bibnamefont {Li}}, \bibinfo {author} {\bibfnamefont
  {C.}~\bibnamefont {Tzschaschel}}, \bibinfo {author} {\bibfnamefont
  {T.}~\bibnamefont {Dinh}}, \bibinfo {author} {\bibfnamefont {Z.}~\bibnamefont
  {Sun}}, \bibinfo {author} {\bibfnamefont {S.-C.}\ \bibnamefont {Ho}},
  \bibinfo {author} {\bibfnamefont {S.-W.}\ \bibnamefont {Lien}}, \bibinfo
  {author} {\bibfnamefont {B.}~\bibnamefont {Singh}}, \bibinfo {author}
  {\bibfnamefont {K.}~\bibnamefont {Watanabe}}, \bibinfo {author}
  {\bibfnamefont {T.}~\bibnamefont {Taniguchi}}, \bibinfo {author}
  {\bibfnamefont {D.~C.}\ \bibnamefont {Bell}}, \bibinfo {author}
  {\bibfnamefont {H.}~\bibnamefont {Lin}}, \bibinfo {author} {\bibfnamefont
  {T.-R.}\ \bibnamefont {Chang}}, \bibinfo {author} {\bibfnamefont {C.~R.}\
  \bibnamefont {Du}}, \bibinfo {author} {\bibfnamefont {A.}~\bibnamefont
  {Bansil}}, \bibinfo {author} {\bibfnamefont {L.}~\bibnamefont {Fu}}, \bibinfo
  {author} {\bibfnamefont {N.}~\bibnamefont {Ni}}, \bibinfo {author}
  {\bibfnamefont {P.~P.}\ \bibnamefont {Orth}}, \bibinfo {author}
  {\bibfnamefont {Q.}~\bibnamefont {Ma}},\ and\ \bibinfo {author}
  {\bibfnamefont {S.-Y.}\ \bibnamefont {Xu}},\ }\bibfield  {title} {\bibinfo
  {title} \textit{Quantum metric nonlinear {Hall} effect in a topological
  antiferromagnetic heterostructure},\ }\href@noop {} {\bibfield  {journal}
  {\bibinfo  {journal} {Science}\ }\textbf {\bibinfo {volume} {381}},\ \bibinfo
  {pages} {181} (\bibinfo {year} {2023})}\BibitemShut {NoStop}%
\bibitem [{\citenamefont {Wang}\ \emph {et~al.}(2023)\citenamefont {Wang},
  \citenamefont {Kaplan}, \citenamefont {Zhang}, \citenamefont {Holder},
  \citenamefont {Cao}, \citenamefont {Wang}, \citenamefont {Zhou},
  \citenamefont {Zhou}, \citenamefont {Jiang}, \citenamefont {Zhang},
  \citenamefont {Ru}, \citenamefont {Cai}, \citenamefont {Watanabe},
  \citenamefont {Taniguchi}, \citenamefont {Yan},\ and\ \citenamefont
  {Gao}}]{Wang2023}%
  \BibitemOpen
  \bibfield  {author} {\bibinfo {author} {\bibfnamefont {N.}~\bibnamefont
  {Wang}}, \bibinfo {author} {\bibfnamefont {D.}~\bibnamefont {Kaplan}},
  \bibinfo {author} {\bibfnamefont {Z.}~\bibnamefont {Zhang}}, \bibinfo
  {author} {\bibfnamefont {T.}~\bibnamefont {Holder}}, \bibinfo {author}
  {\bibfnamefont {N.}~\bibnamefont {Cao}}, \bibinfo {author} {\bibfnamefont
  {A.}~\bibnamefont {Wang}}, \bibinfo {author} {\bibfnamefont {X.}~\bibnamefont
  {Zhou}}, \bibinfo {author} {\bibfnamefont {F.}~\bibnamefont {Zhou}}, \bibinfo
  {author} {\bibfnamefont {Z.}~\bibnamefont {Jiang}}, \bibinfo {author}
  {\bibfnamefont {C.}~\bibnamefont {Zhang}}, \bibinfo {author} {\bibfnamefont
  {S.}~\bibnamefont {Ru}}, \bibinfo {author} {\bibfnamefont {H.}~\bibnamefont
  {Cai}}, \bibinfo {author} {\bibfnamefont {K.}~\bibnamefont {Watanabe}},
  \bibinfo {author} {\bibfnamefont {T.}~\bibnamefont {Taniguchi}}, \bibinfo
  {author} {\bibfnamefont {B.}~\bibnamefont {Yan}},\ and\ \bibinfo {author}
  {\bibfnamefont {W.}~\bibnamefont {Gao}},\ }\bibfield  {title} {\bibinfo
  {title} \textit{Quantum-metric-induced nonlinear transport in a topological
  antiferromagnet},\ }\href@noop {} {\bibfield  {journal} {\bibinfo  {journal}
  {Nature}\ }\textbf {\bibinfo {volume} {621}},\ \bibinfo {pages} {487}
  (\bibinfo {year} {2023})}\BibitemShut {NoStop}%
\bibitem [{\citenamefont {Zhang}\ \emph {et~al.}(2018)\citenamefont {Zhang},
  \citenamefont {Sun},\ and\ \citenamefont {Yan}}]{Zhang2018}%
  \BibitemOpen
  \bibfield  {author} {\bibinfo {author} {\bibfnamefont {Y.}~\bibnamefont
  {Zhang}}, \bibinfo {author} {\bibfnamefont {Y.}~\bibnamefont {Sun}},\ and\
  \bibinfo {author} {\bibfnamefont {B.}~\bibnamefont {Yan}},\ }\bibfield
  {title} {\bibinfo {title} \textit{Berry curvature dipole in {Weyl} semimetal
  materials: An ab initio study},\ }\href@noop {} {\bibfield  {journal}
  {\bibinfo  {journal} {Phys. Rev. B}\ }\textbf {\bibinfo {volume} {97}},\
  \bibinfo {pages} {041101} (\bibinfo {year} {2018})}\BibitemShut {NoStop}%
\bibitem [{\citenamefont {Du}\ \emph {et~al.}(2018)\citenamefont {Du},
  \citenamefont {Wang}, \citenamefont {Lu},\ and\ \citenamefont
  {Xie}}]{Du2018}%
  \BibitemOpen
  \bibfield  {author} {\bibinfo {author} {\bibfnamefont {Z.~Z.}\ \bibnamefont
  {Du}}, \bibinfo {author} {\bibfnamefont {C.~M.}\ \bibnamefont {Wang}},
  \bibinfo {author} {\bibfnamefont {H.-Z.}\ \bibnamefont {Lu}},\ and\ \bibinfo
  {author} {\bibfnamefont {X.~C.}\ \bibnamefont {Xie}},\ }\bibfield  {title}
  {\bibinfo {title} \textit{Band signatures for strong nonlinear {Hall} effect in
  bilayer {WTe$_{2}$}},\ }\href@noop {} {\bibfield  {journal} {\bibinfo
  {journal} {Phys. Rev. Lett.}\ }\textbf {\bibinfo {volume} {121}},\ \bibinfo
  {pages} {266601} (\bibinfo {year} {2018})}\BibitemShut {NoStop}%
\bibitem [{\citenamefont {Zhang}\ \emph {et~al.}(2022)\citenamefont {Zhang},
  \citenamefont {Liang}, \citenamefont {Kaneko}, \citenamefont {Nagaosa},\ and\
  \citenamefont {Tokura}}]{Zhang2022}%
  \BibitemOpen
  \bibfield  {author} {\bibinfo {author} {\bibfnamefont {C.-L.}\ \bibnamefont
  {Zhang}}, \bibinfo {author} {\bibfnamefont {T.}~\bibnamefont {Liang}},
  \bibinfo {author} {\bibfnamefont {Y.}~\bibnamefont {Kaneko}}, \bibinfo
  {author} {\bibfnamefont {N.}~\bibnamefont {Nagaosa}},\ and\ \bibinfo {author}
  {\bibfnamefont {Y.}~\bibnamefont {Tokura}},\ }\bibfield  {title} {\bibinfo
  {title} \textit{Giant {Berry} curvature dipole density in a ferroelectric {Weyl}
  semimetal},\ }\href@noop {} {\bibfield  {journal} {\bibinfo  {journal} {npj
  Quantum Mater.}\ }\textbf {\bibinfo {volume} {7}},\ \bibinfo {pages} {103}
  (\bibinfo {year} {2022})}\BibitemShut {NoStop}%
\bibitem [{\citenamefont {Torre}\ \emph {et~al.}(2021)\citenamefont {Torre},
  \citenamefont {Seyler}, \citenamefont {Zhao}, \citenamefont {Matteo},
  \citenamefont {Scheurer}, \citenamefont {Li}, \citenamefont {Yu},
  \citenamefont {Greven}, \citenamefont {Sachdev}, \citenamefont {Norman},\
  and\ \citenamefont {Hsieh}}]{Torre2021}%
  \BibitemOpen
  \bibfield  {author} {\bibinfo {author} {\bibfnamefont {A.~d.~l.}\
  \bibnamefont {Torre}}, \bibinfo {author} {\bibfnamefont {K.~L.}\ \bibnamefont
  {Seyler}}, \bibinfo {author} {\bibfnamefont {L.}~\bibnamefont {Zhao}},
  \bibinfo {author} {\bibfnamefont {S.~D.}\ \bibnamefont {Matteo}}, \bibinfo
  {author} {\bibfnamefont {M.~S.}\ \bibnamefont {Scheurer}}, \bibinfo {author}
  {\bibfnamefont {Y.}~\bibnamefont {Li}}, \bibinfo {author} {\bibfnamefont
  {B.}~\bibnamefont {Yu}}, \bibinfo {author} {\bibfnamefont {M.}~\bibnamefont
  {Greven}}, \bibinfo {author} {\bibfnamefont {S.}~\bibnamefont {Sachdev}},
  \bibinfo {author} {\bibfnamefont {M.~R.}\ \bibnamefont {Norman}},\ and\
  \bibinfo {author} {\bibfnamefont {D.}~\bibnamefont {Hsieh}},\ }\bibfield
  {title} {\bibinfo {title} \textit{Mirror symmetry breaking in a model insulating
  cuprate},\ }\href@noop {} {\bibfield  {journal} {\bibinfo  {journal} {Nat.
  Phys.}\ }\textbf {\bibinfo {volume} {17}},\ \bibinfo {pages} {777} (\bibinfo
  {year} {2021})}\BibitemShut {NoStop}%
\bibitem [{\citenamefont {Seyler}\ \emph {et~al.}(2020)\citenamefont {Seyler},
  \citenamefont {de~la Torre}, \citenamefont {Porter}, \citenamefont {Zoghlin},
  \citenamefont {Polski}, \citenamefont {Nguyen}, \citenamefont {Nadj-Perge},
  \citenamefont {Wilson},\ and\ \citenamefont {Hsieh}}]{Seyler2020}%
  \BibitemOpen
  \bibfield  {author} {\bibinfo {author} {\bibfnamefont {K.~L.}\ \bibnamefont
  {Seyler}}, \bibinfo {author} {\bibfnamefont {A.}~\bibnamefont {de~la Torre}},
  \bibinfo {author} {\bibfnamefont {Z.}~\bibnamefont {Porter}}, \bibinfo
  {author} {\bibfnamefont {E.}~\bibnamefont {Zoghlin}}, \bibinfo {author}
  {\bibfnamefont {R.}~\bibnamefont {Polski}}, \bibinfo {author} {\bibfnamefont
  {M.}~\bibnamefont {Nguyen}}, \bibinfo {author} {\bibfnamefont
  {S.}~\bibnamefont {Nadj-Perge}}, \bibinfo {author} {\bibfnamefont {S.~D.}\
  \bibnamefont {Wilson}},\ and\ \bibinfo {author} {\bibfnamefont
  {D.}~\bibnamefont {Hsieh}},\ }\bibfield  {title} {\bibinfo {title}
  \textit{Spin-orbit-enhanced magnetic surface second-harmonic generation in
  {Sr$_{2}$IrO$_{4}$}},\ }\href@noop {} {\bibfield  {journal} {\bibinfo
  {journal} {Phys. Rev. B}\ }\textbf {\bibinfo {volume} {102}},\ \bibinfo
  {pages} {201113} (\bibinfo {year} {2020})}\BibitemShut {NoStop}%
\bibitem [{\citenamefont {Hwang}\ \emph {et~al.}(2012)\citenamefont {Hwang},
  \citenamefont {Iwasa}, \citenamefont {Kawasaki}, \citenamefont {Keimer},
  \citenamefont {Nagaosa},\ and\ \citenamefont {Tokura}}]{Hwang2012}%
  \BibitemOpen
  \bibfield  {author} {\bibinfo {author} {\bibfnamefont {H.~Y.}\ \bibnamefont
  {Hwang}}, \bibinfo {author} {\bibfnamefont {Y.}~\bibnamefont {Iwasa}},
  \bibinfo {author} {\bibfnamefont {M.}~\bibnamefont {Kawasaki}}, \bibinfo
  {author} {\bibfnamefont {B.}~\bibnamefont {Keimer}}, \bibinfo {author}
  {\bibfnamefont {N.}~\bibnamefont {Nagaosa}},\ and\ \bibinfo {author}
  {\bibfnamefont {Y.}~\bibnamefont {Tokura}},\ }\bibfield  {title} {\bibinfo
  {title} \textit{Emergent phenomena at oxide interfaces},\ }\href@noop {} {\bibfield
  {journal} {\bibinfo  {journal} {Nat. Mater.}\ }\textbf {\bibinfo {volume}
  {11}},\ \bibinfo {pages} {103} (\bibinfo {year} {2012})}\BibitemShut
  {NoStop}%
\bibitem [{\citenamefont {Pesquera}\ \emph {et~al.}(2012)\citenamefont
  {Pesquera}, \citenamefont {Herranz}, \citenamefont {Barla}, \citenamefont
  {Pellegrin}, \citenamefont {Bondino}, \citenamefont {Magnano}, \citenamefont
  {Sánchez},\ and\ \citenamefont {Fontcuberta}}]{Pesq2012}%
  \BibitemOpen
  \bibfield  {author} {\bibinfo {author} {\bibfnamefont {D.}~\bibnamefont
  {Pesquera}}, \bibinfo {author} {\bibfnamefont {G.}~\bibnamefont {Herranz}},
  \bibinfo {author} {\bibfnamefont {A.}~\bibnamefont {Barla}}, \bibinfo
  {author} {\bibfnamefont {E.}~\bibnamefont {Pellegrin}}, \bibinfo {author}
  {\bibfnamefont {F.}~\bibnamefont {Bondino}}, \bibinfo {author} {\bibfnamefont
  {E.}~\bibnamefont {Magnano}}, \bibinfo {author} {\bibfnamefont
  {F.}~\bibnamefont {Sánchez}},\ and\ \bibinfo {author} {\bibfnamefont
  {J.}~\bibnamefont {Fontcuberta}},\ }\bibfield  {title} {\bibinfo {title}
  \textit{Surface symmetry-breaking and strain effects on orbital occupancy in
  transition metal perovskite epitaxial films},\ }\href@noop {} {\bibfield
  {journal} {\bibinfo  {journal} {Nat. Commun.}\ }\textbf {\bibinfo {volume}
  {3}},\ \bibinfo {pages} {1189} (\bibinfo {year} {2012})}\BibitemShut
  {NoStop}%
\bibitem [{\citenamefont {Sohn}\ \emph {et~al.}(2021)\citenamefont {Sohn},
  \citenamefont {Lee}, \citenamefont {Park}, \citenamefont {Kyung},
  \citenamefont {Hwang}, \citenamefont {Denlinger}, \citenamefont {Kim},
  \citenamefont {Kim}, \citenamefont {Kim}, \citenamefont {Ryu}, \citenamefont
  {Huh}, \citenamefont {Oh}, \citenamefont {Jung}, \citenamefont {Oh},
  \citenamefont {Kim}, \citenamefont {Han}, \citenamefont {Noh}, \citenamefont
  {Yang},\ and\ \citenamefont {Kim}}]{Sohn2021}%
  \BibitemOpen
  \bibfield  {author} {\bibinfo {author} {\bibfnamefont {B.}~\bibnamefont
  {Sohn}}, \bibinfo {author} {\bibfnamefont {E.}~\bibnamefont {Lee}}, \bibinfo
  {author} {\bibfnamefont {S.~Y.}\ \bibnamefont {Park}}, \bibinfo {author}
  {\bibfnamefont {W.}~\bibnamefont {Kyung}}, \bibinfo {author} {\bibfnamefont
  {J.}~\bibnamefont {Hwang}}, \bibinfo {author} {\bibfnamefont {J.~D.}\
  \bibnamefont {Denlinger}}, \bibinfo {author} {\bibfnamefont {M.}~\bibnamefont
  {Kim}}, \bibinfo {author} {\bibfnamefont {D.}~\bibnamefont {Kim}}, \bibinfo
  {author} {\bibfnamefont {B.}~\bibnamefont {Kim}}, \bibinfo {author}
  {\bibfnamefont {H.}~\bibnamefont {Ryu}}, \bibinfo {author} {\bibfnamefont
  {S.}~\bibnamefont {Huh}}, \bibinfo {author} {\bibfnamefont {J.~S.}\
  \bibnamefont {Oh}}, \bibinfo {author} {\bibfnamefont {J.~K.}\ \bibnamefont
  {Jung}}, \bibinfo {author} {\bibfnamefont {D.}~\bibnamefont {Oh}}, \bibinfo
  {author} {\bibfnamefont {Y.}~\bibnamefont {Kim}}, \bibinfo {author}
  {\bibfnamefont {M.}~\bibnamefont {Han}}, \bibinfo {author} {\bibfnamefont
  {T.~W.}\ \bibnamefont {Noh}}, \bibinfo {author} {\bibfnamefont {B.-J.}\
  \bibnamefont {Yang}},\ and\ \bibinfo {author} {\bibfnamefont
  {C.}~\bibnamefont {Kim}},\ }\bibfield  {title} {\bibinfo {title}
  \textit{Sign-tunable anomalous {Hall} effect induced by two-dimensional
  symmetry-protected nodal structures in ferromagnetic perovskite thin films},\
  }\href@noop {} {\bibfield  {journal} {\bibinfo  {journal} {Nat. Mater.}\
  }\textbf {\bibinfo {volume} {20}},\ \bibinfo {pages} {1643} (\bibinfo {year}
  {2021})}\BibitemShut {NoStop}%
\bibitem [{\citenamefont {Train}\ \emph {et~al.}(2009)\citenamefont {Train},
  \citenamefont {Nuida}, \citenamefont {Gheorghe}, \citenamefont {Gruselle},\
  and\ \citenamefont {Ohkoshi}}]{mSHG1}%
  \BibitemOpen
  \bibfield  {author} {\bibinfo {author} {\bibfnamefont {C.}~\bibnamefont
  {Train}}, \bibinfo {author} {\bibfnamefont {T.}~\bibnamefont {Nuida}},
  \bibinfo {author} {\bibfnamefont {R.}~\bibnamefont {Gheorghe}}, \bibinfo
  {author} {\bibfnamefont {M.}~\bibnamefont {Gruselle}},\ and\ \bibinfo
  {author} {\bibfnamefont {S.-i.}\ \bibnamefont {Ohkoshi}},\ }\bibfield
  {title} {\bibinfo {title} \textit{Large magnetization-induced second harmonic
  generation in an enantiopure chiral magnet},\ }\href@noop {} {\bibfield
  {journal} {\bibinfo  {journal} {J. Am. Chem. Soc.}\ }\textbf {\bibinfo
  {volume} {131}},\ \bibinfo {pages} {16838} (\bibinfo {year}
  {2009})}\BibitemShut {NoStop}%
\bibitem [{\citenamefont {Sun}\ \emph {et~al.}(2019)\citenamefont {Sun},
  \citenamefont {Yi}, \citenamefont {Song}, \citenamefont {Clark},
  \citenamefont {Huang}, \citenamefont {Shan}, \citenamefont {Wu},
  \citenamefont {Huang}, \citenamefont {Gao}, \citenamefont {Chen},
  \citenamefont {McGuire}, \citenamefont {Cao}, \citenamefont {Xiao},
  \citenamefont {Liu}, \citenamefont {Yao}, \citenamefont {Xu},\ and\
  \citenamefont {Wu}}]{mSHG2}%
  \BibitemOpen
  \bibfield  {author} {\bibinfo {author} {\bibfnamefont {Z.}~\bibnamefont
  {Sun}}, \bibinfo {author} {\bibfnamefont {Y.}~\bibnamefont {Yi}}, \bibinfo
  {author} {\bibfnamefont {T.}~\bibnamefont {Song}}, \bibinfo {author}
  {\bibfnamefont {G.}~\bibnamefont {Clark}}, \bibinfo {author} {\bibfnamefont
  {B.}~\bibnamefont {Huang}}, \bibinfo {author} {\bibfnamefont
  {Y.}~\bibnamefont {Shan}}, \bibinfo {author} {\bibfnamefont {S.}~\bibnamefont
  {Wu}}, \bibinfo {author} {\bibfnamefont {D.}~\bibnamefont {Huang}}, \bibinfo
  {author} {\bibfnamefont {C.}~\bibnamefont {Gao}}, \bibinfo {author}
  {\bibfnamefont {Z.}~\bibnamefont {Chen}}, \bibinfo {author} {\bibfnamefont
  {M.}~\bibnamefont {McGuire}}, \bibinfo {author} {\bibfnamefont
  {T.}~\bibnamefont {Cao}}, \bibinfo {author} {\bibfnamefont {D.}~\bibnamefont
  {Xiao}}, \bibinfo {author} {\bibfnamefont {W.-T.}\ \bibnamefont {Liu}},
  \bibinfo {author} {\bibfnamefont {W.}~\bibnamefont {Yao}}, \bibinfo {author}
  {\bibfnamefont {X.}~\bibnamefont {Xu}},\ and\ \bibinfo {author}
  {\bibfnamefont {S.}~\bibnamefont {Wu}},\ }\bibfield  {title} {\bibinfo
  {title} \textit{Giant nonreciprocal second-harmonic generation from
  antiferromagnetic bilayer {CrI$_{3}$}},\ }\href@noop {} {\bibfield  {journal}
  {\bibinfo  {journal} {Nature}\ }\textbf {\bibinfo {volume} {572}},\ \bibinfo
  {pages} {497} (\bibinfo {year} {2019})}\BibitemShut {NoStop}%
  \bibitem [{\citenamefont {Roh}\ \emph {et~al.}(2021)\citenamefont {Roh},
  \citenamefont {Kim}, \citenamefont {Park}, \citenamefont {Shin},
  \citenamefont {Yang}, \citenamefont {Noh},\ and\ \citenamefont
  {Lee}}]{Roh2021}%
  \BibitemOpen
  \bibfield  {author} {\bibinfo {author} {\bibfnamefont {C.~J.}\ \bibnamefont
  {Roh}}, \bibinfo {author} {\bibfnamefont {J.~R.}\ \bibnamefont {Kim}},
  \bibinfo {author} {\bibfnamefont {S.}~\bibnamefont {Park}}, \bibinfo {author}
  {\bibfnamefont {Y.~J.}\ \bibnamefont {Shin}}, \bibinfo {author}
  {\bibfnamefont {B.-J.}\ \bibnamefont {Yang}}, \bibinfo {author}
  {\bibfnamefont {T.~W.}\ \bibnamefont {Noh}},\ and\ \bibinfo {author}
  {\bibfnamefont {J.~S.}\ \bibnamefont {Lee}},\ }\bibfield  {title} {\bibinfo
  {title} \textit{Structural symmetry evolution in surface and interface of
  {SrRuO$_{3}$} thin films},\ }\href@noop {} {\bibfield  {journal} {\bibinfo
  {journal} {Appl. Surf. Sci.}\ }\textbf {\bibinfo {volume} {553}},\ \bibinfo
  {pages} {149574} (\bibinfo {year} {2021})}\BibitemShut {NoStop}%
\bibitem [{\citenamefont {Harter}\ \emph {et~al.}(2015)\citenamefont {Harter},
  \citenamefont {Niu}, \citenamefont {Woss},\ and\ \citenamefont
  {Hsieh}}]{Harter2015}%
  \BibitemOpen
  \bibfield  {author} {\bibinfo {author} {\bibfnamefont {J.~W.}\ \bibnamefont
  {Harter}}, \bibinfo {author} {\bibfnamefont {L.}~\bibnamefont {Niu}},
  \bibinfo {author} {\bibfnamefont {A.~J.}\ \bibnamefont {Woss}},\ and\
  \bibinfo {author} {\bibfnamefont {D.}~\bibnamefont {Hsieh}},\ }\bibfield
  {title} {\bibinfo {title} \textit{High-speed measurement of rotational anisotropy
  nonlinear optical harmonic generation using position-sensitive detection},\
  }\href@noop {} {\bibfield  {journal} {\bibinfo  {journal} {Opt. Lett.}\
  }\textbf {\bibinfo {volume} {40}},\ \bibinfo {pages} {4671} (\bibinfo {year}
  {2015})}\BibitemShut {NoStop}%
\bibitem [{\citenamefont {Kresse}\ and\ \citenamefont
  {Joubert}(1999)}]{Kresse}%
  \BibitemOpen
  \bibfield  {author} {\bibinfo {author} {\bibfnamefont {G.}~\bibnamefont
  {Kresse}}\ and\ \bibinfo {author} {\bibfnamefont {D.}~\bibnamefont
  {Joubert}},\ }\bibfield  {title} {\bibinfo {title} \textit{From ultrasoft
  pseudopotentials to the projector augmented-wave method},\ }\href@noop {}
  {\bibfield  {journal} {\bibinfo  {journal} {Phys. Rev. B}\ }\textbf {\bibinfo
  {volume} {59}},\ \bibinfo {pages} {1758} (\bibinfo {year}
  {1999})}\BibitemShut {NoStop}%
\bibitem [{\citenamefont {Kresse}\ and\ \citenamefont
  {Furthm\"uller}(1996)}]{vasp}%
  \BibitemOpen
  \bibfield  {author} {\bibinfo {author} {\bibfnamefont {G.}~\bibnamefont
  {Kresse}}\ and\ \bibinfo {author} {\bibfnamefont {J.}~\bibnamefont
  {Furthm\"uller}},\ }\bibfield  {title} {\bibinfo {title} \textit{Efficient iterative
  schemes for ab initio total-energy calculations using a plane-wave basis
  set},\ }\href@noop {} {\bibfield  {journal} {\bibinfo  {journal} {Phys. Rev.
  B}\ }\textbf {\bibinfo {volume} {54}},\ \bibinfo {pages} {11169} (\bibinfo
  {year} {1996})}\BibitemShut {NoStop}%
\bibitem [{\citenamefont {Perdew}\ \emph {et~al.}(1996)\citenamefont {Perdew},
  \citenamefont {Burke},\ and\ \citenamefont {Ernzerhof}}]{PBE}%
  \BibitemOpen
  \bibfield  {author} {\bibinfo {author} {\bibfnamefont {J.~P.}\ \bibnamefont
  {Perdew}}, \bibinfo {author} {\bibfnamefont {K.}~\bibnamefont {Burke}},\ and\
  \bibinfo {author} {\bibfnamefont {M.}~\bibnamefont {Ernzerhof}},\ }\bibfield
  {title} {\bibinfo {title} \textit{Generalized gradient approximation made simple},\
  }\href@noop {} {\bibfield  {journal} {\bibinfo  {journal} {Phys. Rev. Lett.}\
  }\textbf {\bibinfo {volume} {77}},\ \bibinfo {pages} {3865} (\bibinfo {year}
  {1996})}\BibitemShut {NoStop}%
\bibitem [{\citenamefont {Liechtenstein}\ \emph {et~al.}(1995)\citenamefont
  {Liechtenstein}, \citenamefont {Anisimov},\ and\ \citenamefont
  {Zaanen}}]{Liechtenstein}%
  \BibitemOpen
  \bibfield  {author} {\bibinfo {author} {\bibfnamefont {A.~I.}\ \bibnamefont
  {Liechtenstein}}, \bibinfo {author} {\bibfnamefont {V.~I.}\ \bibnamefont
  {Anisimov}},\ and\ \bibinfo {author} {\bibfnamefont {J.}~\bibnamefont
  {Zaanen}},\ }\bibfield  {title} {\bibinfo {title} \textit{Density-functional theory
  and strong interactions: Orbital ordering in mott-hubbard insulators},\
  }\href@noop {} {\bibfield  {journal} {\bibinfo  {journal} {Phys. Rev. B}\
  }\textbf {\bibinfo {volume} {52}},\ \bibinfo {pages} {R5467} (\bibinfo {year}
  {1995})}\BibitemShut {NoStop}%
\bibitem [{\citenamefont {Marzari}\ and\ \citenamefont
  {Vanderbilt}(1997)}]{Marzari}%
  \BibitemOpen
  \bibfield  {author} {\bibinfo {author} {\bibfnamefont {N.}~\bibnamefont
  {Marzari}}\ and\ \bibinfo {author} {\bibfnamefont {D.}~\bibnamefont
  {Vanderbilt}},\ }\bibfield  {title} {\bibinfo {title} \textit{Maximally localized
  generalized wannier functions for composite energy bands},\ }\href@noop {}
  {\bibfield  {journal} {\bibinfo  {journal} {Phys. Rev. B}\ }\textbf {\bibinfo
  {volume} {56}},\ \bibinfo {pages} {12847} (\bibinfo {year}
  {1997})}\BibitemShut {NoStop}%
\bibitem [{\citenamefont {Mostofi}\ \emph {et~al.}(2014)\citenamefont
  {Mostofi}, \citenamefont {Yates}, \citenamefont {Pizzi}, \citenamefont {Lee},
  \citenamefont {Souza}, \citenamefont {Vanderbilt},\ and\ \citenamefont
  {Marzari}}]{Mostofi}%
  \BibitemOpen
  \bibfield  {author} {\bibinfo {author} {\bibfnamefont {A.~A.}\ \bibnamefont
  {Mostofi}}, \bibinfo {author} {\bibfnamefont {J.~R.}\ \bibnamefont {Yates}},
  \bibinfo {author} {\bibfnamefont {G.}~\bibnamefont {Pizzi}}, \bibinfo
  {author} {\bibfnamefont {Y.-S.}\ \bibnamefont {Lee}}, \bibinfo {author}
  {\bibfnamefont {I.}~\bibnamefont {Souza}}, \bibinfo {author} {\bibfnamefont
  {D.}~\bibnamefont {Vanderbilt}},\ and\ \bibinfo {author} {\bibfnamefont
  {N.}~\bibnamefont {Marzari}},\ }\bibfield  {title} {\bibinfo {title} \textit{An
  updated version of wannier90: A tool for obtaining maximally-localised
  wannier functions},\ }\href@noop {} {\bibfield  {journal} {\bibinfo
  {journal} {Comput. Phys. Commun.}\ }\textbf {\bibinfo {volume} {185}},\
  \bibinfo {pages} {2309} (\bibinfo {year} {2014})}\BibitemShut {NoStop}%
\bibitem [{\citenamefont {Franchini}\ \emph {et~al.}(2012)\citenamefont
  {Franchini}, \citenamefont {Kováčik}, \citenamefont {Marsman},
  \citenamefont {Murthy}, \citenamefont {He}, \citenamefont {Ederer},\ and\
  \citenamefont {Kresse}}]{Franchini}%
  \BibitemOpen
  \bibfield  {author} {\bibinfo {author} {\bibfnamefont {C.}~\bibnamefont
  {Franchini}}, \bibinfo {author} {\bibfnamefont {R.}~\bibnamefont
  {Kováčik}}, \bibinfo {author} {\bibfnamefont {M.}~\bibnamefont {Marsman}},
  \bibinfo {author} {\bibfnamefont {S.~S.}\ \bibnamefont {Murthy}}, \bibinfo
  {author} {\bibfnamefont {J.}~\bibnamefont {He}}, \bibinfo {author}
  {\bibfnamefont {C.}~\bibnamefont {Ederer}},\ and\ \bibinfo {author}
  {\bibfnamefont {G.}~\bibnamefont {Kresse}},\ }\bibfield  {title} {\bibinfo
  {title} \textit{Maximally localized {Wannier} functions in {LaMnO$_3$} within {PBE +
  U}, hybrid functionals and partially self-consistent {GW}: an efficient route
  to construct ab initio tight-binding parameters for e$_{\rm g}$
  perovskites},\ }\href@noop {} {\bibfield  {journal} {\bibinfo  {journal}
  {Journal of Physics: Condensed Matter}\ }\textbf {\bibinfo {volume} {24}},\
  \bibinfo {pages} {235602} (\bibinfo {year} {2012})}\BibitemShut {NoStop}%
\bibitem [{\citenamefont {Wu}\ \emph {et~al.}(2018)\citenamefont {Wu},
  \citenamefont {Zhang}, \citenamefont {Song}, \citenamefont {Troyer},\ and\
  \citenamefont {Soluyanov}}]{QuanSheng}%
  \BibitemOpen
  \bibfield  {author} {\bibinfo {author} {\bibfnamefont {Q.}~\bibnamefont
  {Wu}}, \bibinfo {author} {\bibfnamefont {S.}~\bibnamefont {Zhang}}, \bibinfo
  {author} {\bibfnamefont {H.-F.}\ \bibnamefont {Song}}, \bibinfo {author}
  {\bibfnamefont {M.}~\bibnamefont {Troyer}},\ and\ \bibinfo {author}
  {\bibfnamefont {A.~A.}\ \bibnamefont {Soluyanov}},\ }\bibfield  {title}
  {\bibinfo {title} \textit{Wanniertools: An open-source software package for novel
  topological materials},\ }\href@noop {} {\bibfield  {journal} {\bibinfo
  {journal} {Comput. Phys. Commun.}\ }\textbf {\bibinfo {volume} {224}},\
  \bibinfo {pages} {405} (\bibinfo {year} {2018})}\BibitemShut {NoStop}%
\bibitem [{\citenamefont {Stokes}\ and\ \citenamefont {Hatch}(2005)}]{Stokes}%
  \BibitemOpen
  \bibfield  {author} {\bibinfo {author} {\bibfnamefont {H.~T.}\ \bibnamefont
  {Stokes}}\ and\ \bibinfo {author} {\bibfnamefont {D.~M.}\ \bibnamefont
  {Hatch}},\ }\bibfield  {title} {\bibinfo {title} \textit{{{\it FINDSYM}: program for
  identifying the space-group symmetry of a crystal}},\ }\href
  {https://doi.org/10.1107/S0021889804031528} {\bibfield  {journal} {\bibinfo
  {journal} {Journal of Applied Crystallography}\ }\textbf {\bibinfo {volume}
  {38}},\ \bibinfo {pages} {237} (\bibinfo {year} {2005})}\BibitemShut
  {NoStop}%
  \bibitem{Yan2024}
  D. Kaplan, T. Holder, and B. Yan, \textit{Unification of nonlinear anomalous Hall effect and nonreciprocal magnetoresistance in metals by the quantum geometry}, Phys. Rev. Lett.
\textbf{132,} 026301 (2024).
  
  
\end{thebibliography}

%

\clearpage

\newpage

\end{document}